\documentclass[usenatbib]{mn2e}
\usepackage{amsmath, amssymb}
\usepackage{amsfonts}
\usepackage{epsfig,rotating}

\def\simeq{
\mathrel{\raise.3ex\hbox{$\sim$}\mkern-14mu\lower0.4ex\hbox{$-$}}
}

\def\ltsima{$\; \buildrel < \over \sim \;$}
\def\simlt{\lower.5ex\hbox{\ltsima}}
\def\gtsima{$\; \buildrel > \over \sim \;$}
\def\simgt{\lower.5ex\hbox{\gtsima}}

\def\lsun{{\rm L_{\odot}}}
\def\msun{{\rm M_{\odot}}}
 
\def\lta{\la}

\def\be{\begin{equation}}
\def\ee{\end{equation}}

\def\del#1{{}}
\def\ltsima{$\; \buildrel < \over \sim \;$}
\def\simlt{\lower.5ex\hbox{\ltsima}}
\def\gtsima{$\; \buildrel > \over \sim \;$}
\def\simgt{\lower.5ex\hbox{\gtsima}}
\def\sgra{Sgr~A$^*$}

\newcommand{\apj}{ApJ}

\newcommand{\mnras}{MNRAS}
\newcommand{\aap}{A\&A}
\newcommand{\araa}{ARA\&A}
\newcommand{\apjl}{ApJL}
\newcommand{\aj}{AJ}
\newcommand{\nat}{Nature}

\newcommand{\pasj}{PASJ}

\title[\sgra\ shapes the Central Molecular Zone]{AGN activity and nuclear starbursts: \sgra\ activity shapes the Central Molecular Zone}
\author[Kastytis~Zubovas]{Kastytis~Zubovas\\ 
Center for Physical Sciences and Technology, Savanori\c{u} 231, Vilnius LT-02300, Lithuania\\
{E-mail:~} {\rm kastytis.zubovas@ftmc.lt}}

\begin{document}

\maketitle

\begin{abstract}

The Central Molecular Zone (CMZ) of the Milky Way shows several
peculiar properties: a large star formation rate, some of the most
massive young star clusters and molecular clouds in the Galaxy, and a
twisted ring morphology in molecular gas. In this paper, I use SPH
simulations to show that most of these properties can be explained as
due to a recent outburst of AGN activity in \sgra, the central
supermassive black hole of the Milky Way. In particular, the narrow
ring of dense gas, massive gas clouds, young star clusters and an
elevated SFR can all be caused by the passage of an AGN outflow
through the system, which compresses the gas and triggers
fragmentation. Furthermore, I show that the asymmetric distribution of
gas, as observed in the CMZ, can be produced by outflow-induced
instabilities from an initially axisymmetric gas disc. Angular
momentum mixing in the disc produces some low angular momentum
material, which can subsequently feed \sgra. These processes can occur
in any galaxy that experiences an AGN episode, leading to bursts of
nuclear star formation much stronger than pure bar-driven mass inflows
would predict.

\end{abstract}


\section{Introduction}

Over the past two decades, various observations revealed that most
galaxies harbour supermassive black holes (SMBHs) at their centres
\citep{Kormendy1995ARA&A, Magorrian1998AJ}. It is expected that these
SMBHs and their immediate surroundings spend a few percent of their
lifetimes as active galactic nuclei (AGNs), accreting gas at rates
exceeding a few percent of their Eddington limit $\dot{M}_{\rm Edd} =
2.2\times10^{-8}\left(M_{\rm BH}/\msun\right) \msun$~yr$^{-1}$. During
this time, the AGN luminosity can exceed that of the whole host
galaxy. The AGN radiation can heat the surrounding gas by
photoionization and Compton heating. More importantly,
quasi-relativistic winds \citep{Tombesi2010ApJ, Tombesi2014MNRAS} and
massive molecular outflows \citep{Sturm2011ApJ, Cicone2014A&A,
  Cicone2015A&A} have been detected in AGN, the latter most likely
driven by the former \citep{Tombesi2015arXiv}. The observed outflow
energy rate is $\sim 5\%$ of the AGN luminosity, as predicted by
analytical \citep{Zubovas2012ApJ} and numerical \citep{Costa2014MNRAS,
  Costa2015MNRAS} models. Such outflows affect the AGN host galaxy on
scales ranging from the immediate AGN environment to the
circumgalactic medium. Importantly, footprints of these effects can
remain visible for a much longer time than the duration of the AGN
phase itself, allowing us to probe the past activity of presently
quiescent galactic nuclei.

A prototypical and well-studied quiescent galactic nucleus is our own
Galctic Centre (GC), which contains a radio source \sgra, identified
as an $M_{\rm BH} = 4.3 \times 10^6 \; \msun$ SMBH \citep{Ghez2008ApJ,
  Gillessen2009ApJ}. The current bolometric luminosity of \sgra\ is
very low, $L_{\rm bol} \sim 300 \lsun \sim 10^{-9} L_{\rm Edd}$
\citep[e.g.,][]{Melia2001ARA&A}. There is evidence that this low
luminosity is caused by two effects: there is little gas supplied into
the several-pc wide sphere of influence around \sgra
\citep{Bower2003ApJ, Marrone2006ApJ}, and the gas that is supplied is
accreted with very low efficiency \citep{Wang2013Sci}.

However, there is growing evidence that \sgra\ used to be much brighter
in the past. Light echoes from molecular clouds in the Galactic centre
reveal that \sgra\ X-ray luminosity was $L_{\rm X} \sim 10^{39}$~erg
s$^{-1}$ about$ 100$ years ago \citep{Revnivtsev2004A&A,
  Terrier2010ApJ} and perhaps on several other occasions in the past
$\sim 10^3$ years \citep{Ryu2012arXiv}. On longer timescales, there is
evidence of significant activity some time during the past 10
Myr. This evidence comes from the presence of young stars in the
central parsec \citep{Paumard2006ApJ, Hobbs2009MNRAS} and the two
large gamma-ray emitting bubbles \citep[the {\it Fermi}
  bubbles;][]{Su2010ApJ}, which have properties consistent with being
inflated several Myr ago by either an AGN jet \citep{Guo2012ApJ} or
wind-driven outflow \citep[also see Section \ref{sec:analytic}
  below]{Zubovas2011MNRAS, Zubovas2012MNRASa}. In particular, in these
two papers my colleagues and I showed that an Eddington-limited AGN
outburst that started 6 Myr ago and lasted for 1 Myr would have
produced two bubbles with sizes, shapes and energy content consistent
with observations.

Given this evidence of increased past activity, it is interesting to
check whether footprints of that activity might be present closer to
the GC, such as in the Central Molecular Zone (CMZ) of the Galaxy. The
CMZ is a ring-like structure of predominantly molecular gas
\citep{Morris1996ARA&A, Jones2011MNRAS, Molinari2011ApJ}. It is
oriented parallel to the plane of the Galaxy, with a radius $R_{\rm
  CMZ} \sim 200$~pc and thickness $h_{\rm CMZ} \sim 100$~pc
\citep{Morris1996ARA&A}, giving an approximate aspect ratio $H/R
\simeq 0.25$. The total gas mass contained in this region is $M_{\rm
  CMZ} \sim 10^8 \; \msun$ \citep{Morris1996ARA&A,
  Pierce-Price2000ApJ, Molinari2011ApJ}.

The CMZ contains several star formation regions, such as the molecular
cloud Sgr B2 \citep[the most massive GMC in the Galaxy; $M_{\rm Sgr
    B2} \simeq 2 \times 10^6 \; \msun$;][]{Revnivtsev2004A&A}, Sgr C
and Sgr D \citep{Pierce-Price2000ApJ}. Two young massive star
clusters, Arches \citep[$t_{\rm age} \simeq 3\pm 1$~Myr, $M \simeq
  4-7\times 10^4 \msun$,][]{Martins2008A&A, PortegiesZwart2002ApJ,
  Figer2002ApJ} and Quintuplet \citep[$t_{\rm age} \simeq 4\pm 1$~Myr,
  $M \simeq 10^4 \msun$,][]{Figer1999ApJb} are also associated with
the region. The mean star formation rate in the region, $\dot{M}_*
\sim 0.06-0.08 \; \msun$~yr$^{-1}$ \citep{Yusef-Zadeh2009ApJ,
  Immer2012A&A, Longmore2013MNRAS}, would deplete the molecular gas
reservoir in $t_{\rm dep} \sim 10^9$~yr $\ll t_{\rm Galaxy} \sim
10^{10}$~yr, therefore it seems likely that the present star formation
rate is elevated by some recent process, i.e. one that occurred at
most a few dynamical times ago \citep[see, however,][who show that the
  elevated star formation rate can be maintained by material transport
  via bars from further out in the MW]{Rodriguez2008A&A}.

Recent {\it Herschel} observations reveal a complex structure of
numerous clouds and filaments comprising the CMZ
\citep{Molinari2011ApJ}. They are mostly positioned on a narrow
elliptical ring, which is twisted and offset from the dynamical centre
of the Galaxy. The line-of-sight velocities of the clouds are larger
than would be expected if the gas was moving along the ring. In
particular, the ``20 km/s'' and ``50 km/s'' clouds certainly do not
follow the ring. Therefore the ring is most likely just a transient
structure, created by an accidental accumulation of gas.

In this paper, I present numerical simulations designed to show what
effect a recent burst of AGN activity in \sgra\ might have had upon
the CMZ properties. Starting from idealised initial conditions, the
AGN wind shapes the CMZ into its present-day elliptical configuration
and triggers gravitational instabilities which lead to formation of
massive young star clusters. The overall conclusion is that AGN
activity might be as important a factor in CMZ evolution as the
perturbation due to the bar potential of the Galaxy.

The paper is structured as follows. I briefly review the AGN
wind-driven outflow model and derive the important quantities
regarding its effect upon the cold CMZ gas in section
\ref{sec:analytic}. Then, I describe the numerical simulation setup
(section \ref{sec:nummodel}) and present their results (section
\ref{sec:results}), paying particular attention to the morphology of
the gas and orbits of star clusters, the star formation rate and the
global properties of the star cluster population. I discuss the main
implications of these results in section \ref{sec:discuss} and
summarize them in section \ref{sec:concl}.

\section{AGN winds and their effect on dense gas} \label{sec:analytic}

\subsection{Wind feedback model}

Observations show that a large fraction of AGN have fast ($v_{\rm w}
\sim 0.1 c$) winds with kinetic power $L_{\rm kin} \sim 0.05 L_{\rm
  AGN}$ \citep{Tombesi2010A&A, Tombesi2010ApJ}. The ubiquity of
observed winds suggests that they are emitted with a wide opening
angle, perhaps quasi-spherically \citep{Nardini2015Sci}. These winds
shock against the galactic ISM and drive large-scale outflows
\citep{King2010MNRASa}. In principle, if the shock occurs very close
to the SMBH, the inverse-Compton cooling allows the wind to rapidly
lose most of its energy and the resultant flow is momentum-driven
\citep{King2003ApJ}. However, in practice, the electron-ion
thermalisation timescale in the two-temperature plasma of the shocked
wind is long enough so that cooling is efficient only at distances
$<1$~pc from the AGN \citep{Faucher2012MNRASb} and the outflows are
essentially always energy-driven. Nevertheless, in a non-uniform ISM,
most of the outflow energy leaks out through low-density channels,
effectively recovering the momentum-driven solution for the dense gas
\citep{Nayakshin2014MNRAS, Zubovas2014MNRASb, Gabor2014MNRAS}. Such
outflows are then good candidates for communicating the AGN luminosity
to the host galaxy and establishing the $M-\sigma$ relation
\citep{King2003ApJ, King2010MNRASa}. The ram pressure of the wind can
only push away dense gas if the AGN luminosity is higher than the
critical value $L_{\rm \sigma} \sim 3\times 10^{45}\sigma_{100}^4
$~erg~s$^{-1}$, where $\sigma \equiv 100\sigma_{100}$~km/s is the
velocity dispersion in the host galaxy. Assuming the AGN to be
Eddington-limited, this translates into a minimum mass $M_\sigma \sim
2.3 \times 10^7 \sigma_{100}^4 \; \msun$.

In our Galactic centre, one then expects that any outflow would affect
the CMZ predominantly via the ram pressure of the wind. This pressure
exerts a force
\begin{equation}
\dot{p} = \frac{L}{c} \lta \frac{4 \pi G M_{\rm SMBH}}{\kappa_{\rm e}}
\simeq 1.8 \times 10^{34} \; {\rm dyn},
\end{equation}
where I used the value of \sgra\ mass $M = 4.3\times 10^6 \; \msun$
\citep{Ghez2008ApJ, Gillessen2009ApJ}. In comparison, the weight of
the CMZ is
\begin{equation}
\begin{split}
W_{\rm CMZ} & \simeq \frac{G M_{\rm CMZ} M\left(<R_{\rm
    CMZ}\right)}{R_{\rm CMZ}^2} \simeq \frac{M_{\rm CMZ}
  \sigma^2}{R_{\rm CMZ}} \\ & \simeq 3 \times 10^{34} M_8
\sigma_{100}^2 R_{200}^{-1} \; {\rm dyn},
\end{split}
\end{equation}
We see that even an Eddington-limited outburst from \sgra\ is unable
to lift the CMZ and remove it from the Galaxy. Note, however, that the
difference between the two quantities is only a factor $\sim2$, so
even a modest retention of the outflow energy (rather than just its
momentum) can disperse the CMZ. This may lead to the CMZ being eroded
over a longer timescale, as its surface material absorbs some of the
outflow energy, heats up and evaporates.

\subsection{Mixing of CMZ material}

Even though the CMZ is not dispersed, the AGN wind can significantly
perturb it. The ram pressure is not negligible compared to the weight
of the whole CMZ, so a significant fraction of the CMZ material is
pushed outwards and can mix with the rest.

The CMZ is located outside the sphere of influence of the SMBH, so the
circular velocity of its components is approximately independent of
radius. Hence the specific angular momentum of CMZ gas increases
linearly with radius and spans more than an order of magnitude. When
gas with different angular momentum is mixed together, the result is
an averaging of this quantity \citep{Hobbs2011MNRAS}.  Gas injected
from the centre gains angular momentum, while gas from outside loses
it, in a process similar to a time-inverted viscous ring evolution
(except that the process responsible for the mixing of angular
momentum is dynamical perturbations rather than viscosity). We can
make a rough estimate of the radius at which the CMZ gas would collect
after this perturbation. If we assume the CMZ follows an $R^{-2}$
radial density profile with constant $H/R$, the mass at each radius is
just $M(R \rightarrow R+\Delta R) = M_{\rm CMZ} \Delta R / R_{\rm
  CMZ}$. If the CMZ is affected out to radius $R_{\rm out} < R_{\rm
  CMZ}$ and all gas is moving on circular orbits initially, the total
angular momentum in the affected region is $M_{\rm out} v_{\rm circ}
R_{\rm out} / 2$, therefore the gas circularizes at a mean radius of
$R_{\rm out} / 2$. If the whole CMZ was affected, the circularization
radius is 100 pc. Furthermore, gas that was initially outside and
inside the circularization radius moves inward and outward,
respectively, increasing the density contrast between the gas
collected at the circularization radius and gas outside of it. This
effect can turn an initially radially widespread gas distribution into
a narrow ring.

The background density outside the sphere of influence of \sgra,
assuming a singular isothermal sphere potential, is
\begin{equation}
\rho_{\rm SIS} = \frac{\sigma^2}{2 \pi G R^2} \simeq 2.5 \times 10^{-21}
\sigma_{100}^2 R_{100}^{-2} \; {\rm g} \; {\rm cm}^{-3},
\end{equation}
which, assuming fully ionised gas of Solar metallicity ($\mu = 0.63$),
gives
\begin{equation}
n_{\rm SIS} \simeq 2.5 \times 10^3 \sigma_{100}^2 R_{100}^{-2} \; {\rm
  cm}^{-3}.
\end{equation}
The mean gas density within the CMZ, again assuming an $R^{-2}$
density profile, is $\rho_{\rm CMZ} \sim 10^{-21} M_8 R_{100}^{-2}$~g
cm$^{-3} \simeq 0.4 \rho_{\rm SIS}$. When the perturbed gas begins
circularizing at around $R_{\rm CMZ}/2$, it can easily pass the
$\rho_{\rm SIS}$ threshold and become self-gravitating.

Gravitational instabilities can lead to formation of filaments and
clumps of gas, as well as star clusters. This is a stochastic process,
so I turn to numerical simulations to investigate the properties of
the structures forming in a perturbed CMZ.

\section{Numerical model}\label{sec:nummodel}

\begin{table*}
\begin{tabular}{c | c c | c c c c}
Model ID & $l$ & $t_{\rm q}$ (Myr) & $t_{\rm sink}$ (Myr) & $\left<\dot{M}_{\rm sink}\right>$ ($\msun$ yr$^{-1}$) & $M_{\rm sink, tot}$ ($10^6 \msun$) & $a/b$ \\
\hline
\hline
Control & $-$ & $0$ & $-$ & $0$ & $0$ & $-$ \\
\hline
Base1 & $1$ & $1$ & $4.5$ & $15.0$ & $26.7$ & $1.39$ \\
Base2 & $1$ & $1$ & $4.6$ & $11.3$ & $27.6$ & $1.34$ \\
Base3 & $1$ & $1$ & $-$   & $4.9$  & $29.7$ & $1.53$ \\
Base4 & $1$ & $1$ & $3.4$ & $12.7$ & $37.8$ & $1.39$ \\
Base5 & $1$ & $1$ & $4.0$ & $14.2$ & $37.8$ & $1.43$ \\
\hline
Short1 & $1$ & $0.7$ & $2.6$ & $11.8$ & $28.1$ & $1.43$ \\
Short2 & $1$ & $0.7$ & $3.0$ & $13.6$ & $28.9$ & $1.27$ \\
Short3 & $1$ & $0.7$ & $3.5$ & $13.2$ & $24.3$ & $1.55$ \\
Short4 & $1$ & $0.7$ & $3.0$ & $13.2$ & $27.9$ & $1.38$ \\
\hline
L0.3 & $0.3$ & $1$           & $-$   & $1.7$ & $10.2$ & $1.16$ \\
T0.3 & $1$ & $0.3$           & $1.4$ & $3.7$ & $23.0$ & $1.09$ \\
T1.5 & $1$ & $1.5$           & $-$   & $2.1$ & $12.6$ & $6.78^*$ \\
L0.667T1.5 & $0.667$ & $1.5$ & $-$   & $2.2$ & $12.9$ & $4.23^*$ \\
\hline
\hline

\end{tabular}
\caption{Parameters of the numerical models and most important
  results. The first column shows the model ID. The next two columns
  give the parameters: the AGN Eddington ratio, $l \equiv L/L_{\rm
    Edd}$, and the duration of AGN phase ir Myr, $t_{\rm q}$. The
  final five columns are the primary results: time of onset of
  significant fragmentation (see text), mean fragmentation rate after
  the onset, total sink particle mass at $t = 6$~Myr, CMZ ring
  major-to-minor axis ratio at $t = 6$~Myr. Models Base1 to Base5 and
  models Short1 to Short5 differ in initial microscopic particle
  distribution, but have identical parameters. Numbers labelled with
  $^*$ use a lower density threshold to identify the ring due to lack
  of significant structures.}
\label{table:param}
\end{table*}

I run SPH simulations of the CMZ affected by a burst of AGN
activity. The code used is GADGET-3, an updated version of the
publicly available GADGET-2 \citep{Springel2005MNRAS}. It is a hybrid
N-body/SPH code with individual particle timesteps and adaptive
smoothing.

With these simulations, I intend to show how AGN activity can create
an asymmetric distribution of gas and star clusters from initially
smooth and regular conditions. Therefore, I set up the CMZ as a smooth
disc with total mass $M = 10^8 \; \msun$, extending from an inner
radius $r_{\rm in} = 5$~pc to outer radius $r_{\rm out} = 200$~pc,
with a radial density profile $\rho \propto r^{-2}$. The CMZ has a
constant rotational velocity $v_{\phi} = 141$~km/s, which makes it
rotationally supported against the gravity of the background
potential. The disc gas is initialized with a constant temperature $T
= 3.2\times10^4$~K, giving a constant $H/R = 0.25$, i.e. the scale
height of the disc is $50$~pc from the midplane at the outer edge. In
the vertical direction, the gas density decreases exponentially away
from the midplane. This set up results in a surface density $\Sigma
\simeq 1.5 \times 10^4 \left(r/r_{\rm in}\right)^{-1} \;
\msun$~pc$^{-2}$ and a radius-independent Toomre Q parameter $Q \simeq
2.5$. Thus the disc is initially marginally stable against
axisymmetric perturbations and self-gravity.

The CMZ has $N_{\rm cmz} = 5\times 10^5$ particles, giving a particle
mass $m_{\rm SPH} = 200 \; \msun$ and mass resolution $m_{\rm res} =
40m_{\rm sph} = 8000 \; \msun$. This is fine enough to resolve large
molecular clouds and clusters. The spatial resolution is varying due
to the nature of SPH. In the initial disc, it ranges from $\sim 1$~pc
at the inner edge to $\sim 13$~pc at the outer edge; in the densest
regions, the spatial resolution goes down to the minimum allowed
smoothing length of $0.01$~pc.

The disc is surrounded with a halo of diffuse ($\rho_{\rm g} = 10^{-3}
\rho_{\rm SIS}$) gas extending out to $5$~kpc. Although a very crude
approximation, this density approximately corresponds to the observed
gas density in the Milky Way \citep{Kalberla2008A&A}. The halo gas
temperature is set at $T_{\rm h} = 2.4 \times 10^5$~K, equal to the
virial temperature of the surrounding static, spherically symmetric
isothermal background potential with velocity dispersion $\sigma =
100$~km/s. I comment on the effects of the assumption of spherical
symmetry in Section \ref{sec:current_prop}.

\begin{figure}
  \centering
    \includegraphics[trim = 6mm 23mm 6mm 0, clip, width=0.48 \textwidth]{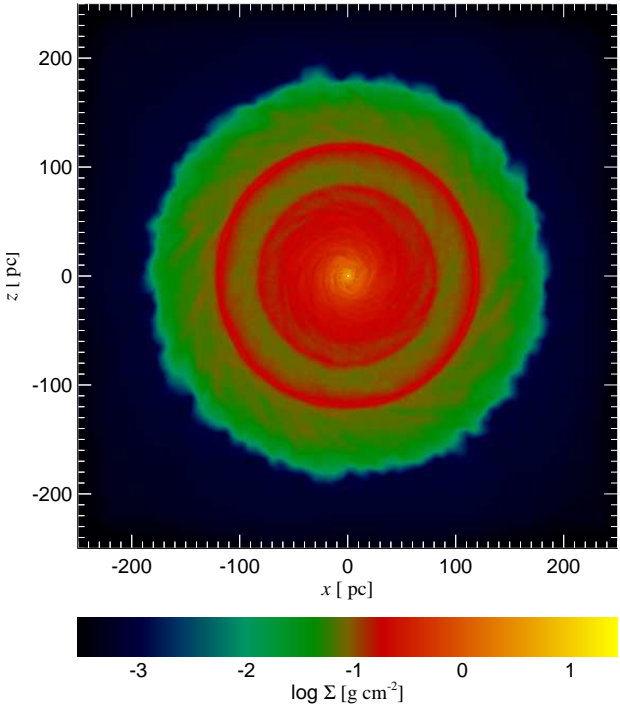}
  \caption{Projected gas density of the Control simulation at $t =
    6$~Myr. Several spiral patterns form in the disc, but there is no
    collapse or fragmentation.}
  \label{fig:control_morph}
\end{figure}

A SMBH with mass $M_{\rm BH} = 4.3 \times 10^6 \; \msun$ is embedded
in the centre of the gas distribution. The sphere of influence of the
SMBH extends only to $R_{\rm infl} \sim 1$~pc, therefore the SMBH
gravity does not affect the dynamics of the gas. At the start of each
simulation, the SMBH is turned on as an AGN of luminosity $L = lL_{\rm
  Edd}$ and radiates at this luminosity for a time $t_{\rm q}$, after
which the luminosity drops to zero. The AGN effect on the gas is
modelled in two ways. First of all, the simulation employs a
heating-cooling function appropriate for optically thin gas of Solar
metallicity exposed to quasar radiation \citep{Sazonov2005MNRAS}, with
a temperature floor of $T_{\rm floor} = 10^4$~K. Secondly, AGN wind
feedback is modelled with the help of virtual particles
\citep{Nayakshin2009MNRAS}. These particles are emitted by the SMBH
every timestep. They have only trace mass, but carry momentum $p_{\rm
  virt} = k m_{\rm SPH} \sigma$ each, where $k = 0.1$ is an arbitrary
constant, and energy $E_{\rm virt} = p_{\rm virt} c / 2$; $p_{\rm
  virt}$ also sets the number of virtual particles emitted each
timestep. They move at a constant velocity $v_{\rm w} = 0.1 c$
radially outward from the source. Whenever an SPH particle contains a
virtual particle within its smoothing kernel, interaction between the
two begins. The virtual particle gives up a fraction of its momentum
and energy to the surrounding SPH particles, weighted by the SPH
particles' smoothing kernel. The fraction given up in each interaction
is set so that it takes $\sim 10$ steps for each virtual particle to
lose $99\%$ of its momentum/energy, at which point it is removed from
the simulation.  This method of transferring feedback
self-consistently accounts for different optical depth of gas in
different directions and enables the formation of anisotropic
feedback-inflated bubbles. In all simulations presented here, the wind
shock is assumed to be adiabatic, therefore all of the wind kinetic
energy $L_{\rm w} = 0.05 L_{\rm AGN}$ is transferred to the gas.

In order to speed up the simulations and for easier tracking of the
star formation process, we convert gas particles into sink particles
according to a Jeans' condition. Whenever a gas particle density
increases above the critical value
\begin{equation} \label{eq:rhojeans}
\rho_{\rm J} = \left(\frac{\pi k_{\rm B} T}{\mu m_{\rm p} G}\right)^3
m_{\rm sph}^{-2} \simeq 1.5 \times 10^{-12} T_4^3 \; {\rm g} \;
{\rm cm}^{-3},
\end{equation}
\begin{equation}
n_{\rm J} \simeq 1.4 \times 10^{12} T_4^3 \; {\rm cm}^{-3},
\end{equation}
where $T_4 \equiv T/10^4$~K, it is converted into a sink particle of
the same mass, which subsequently interacts with the other particles
only via gravity and can accrete other gas or sink particles once they
come within $10^{-4}$~pc. The critical density is such that the
corresponding Jeans' mass is equal to $m_{\rm SPH}$. In all
simulations, most gas particles stay well below the density threshold,
with density increasing significantly only in self-gravitating clumps
on sub-parsec scales; reducing the threshold by several orders of
magnitude does not significantly affect the results.

\begin{figure*}
  \centering
    \includegraphics[trim = 4mm 23mm 8mm 0, clip, width=0.32 \textwidth]{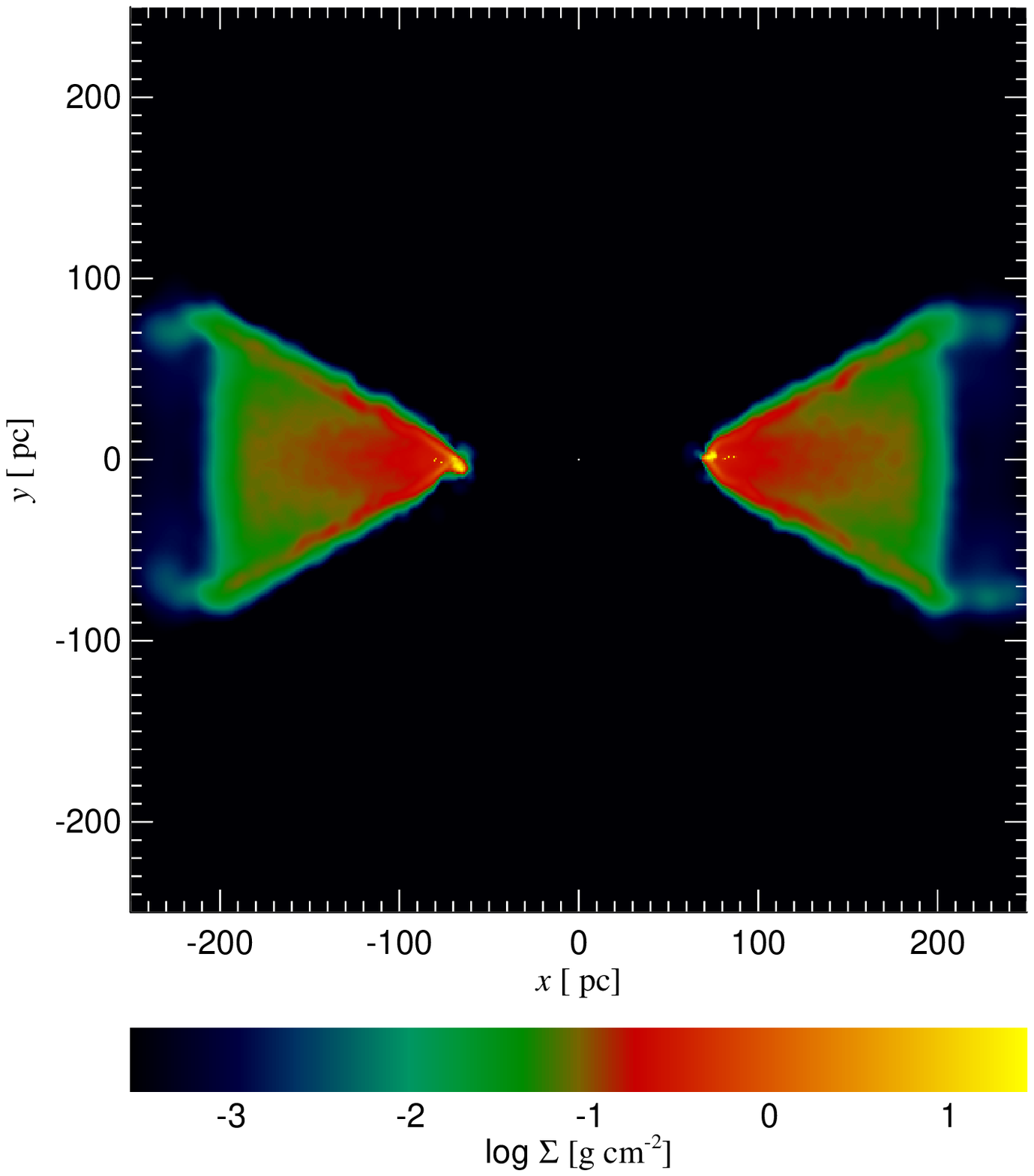}
    \includegraphics[trim = 4mm 23mm 8mm 0, clip, width=0.32 \textwidth]{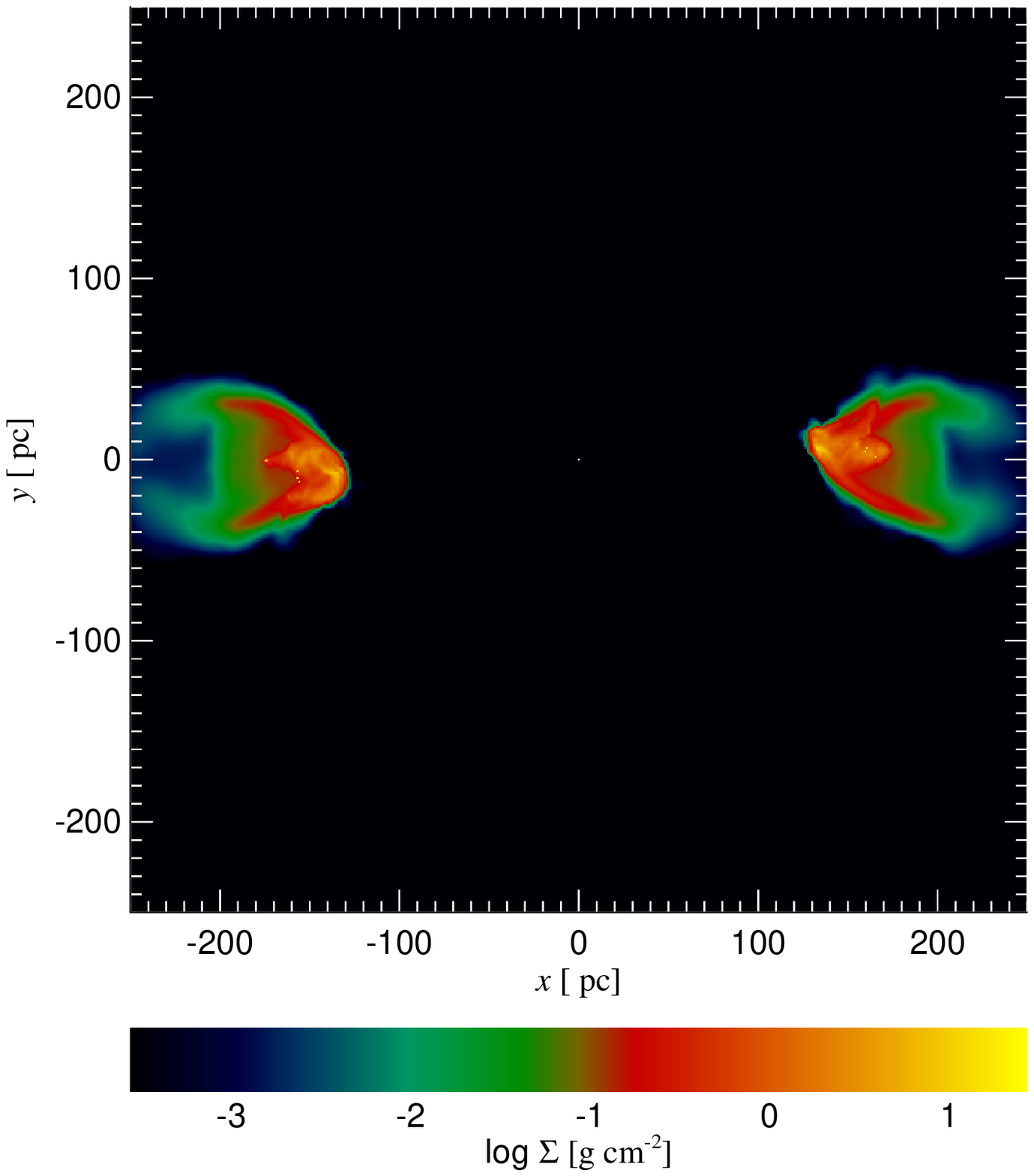}
    \includegraphics[trim = 4mm 23mm 8mm 0, clip, width=0.32 \textwidth]{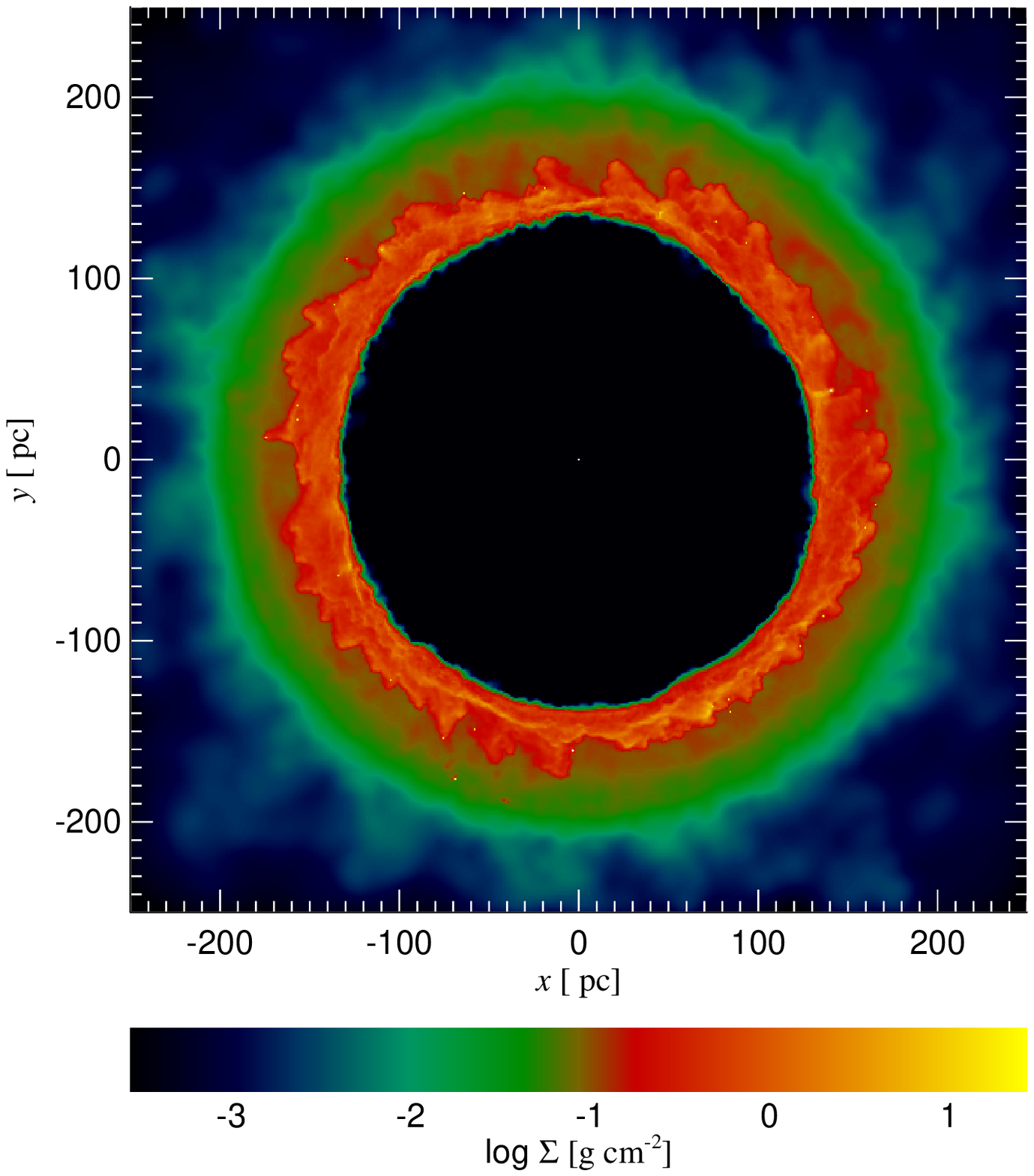}
  \caption{Projected gas density of fiducial run Base1 at early
    times. {\bf Left}: cut side view at $t = 0.5$~Myr, showing the
    radial structure of the CMZ disc. The central regions are pushed
    together and begin collapsing vertically, exposing parts of the
    disc further out to AGN feedback. {\bf Middle}: cut side view at
    $t = t_{\rm q} = 1$~Myr. The whole disc is affected, with dense
    and thick inner ring and surface layers pushing into the outer
    parts. {\bf Right}: top view at $t = t_{\rm q} = 1$~Myr.  A
    transition region is visible, with spiral perturbations and
    self-gravitating clumps.}
  \label{fig:morph_early}
\end{figure*}

There are fourteen simulations in total; their parameters - $l$ and
$t_{\rm q}$ - and main results are given in Table \ref{table:param}. I
run one Control simulation with no AGN activity, in order to check
whether the initial gas distribution is stableto its own self-gravity
without external perturbations. Five simulations (Base1 to Base5) have
identical parameters, but use stochastically different initial
conditions, as the particles comprising the CMZ were cut from
different parts of a relaxed glass distribution. The choice of $l = 1$
and $t_{\rm q} = 1$~Myr is motivated by previous work
\citep{Zubovas2012MNRASa}, where we found that these parameters
produce hot gas cavities very similar to the observed {\it Fermi}
bubbles. Four more simulations (Short1 to Short4) also differ only in
microscopic particle distribution and have $t_{\rm q} = 0.7$~Myr; this
choice is motivated by the results of the Base simulations, as
described below. The final four simulations show the effects of
varying $l$ and $t_{\rm q}$ further, as these parameters are least
constrained by observations and current models. All but one of these
four simulations use the same initial conditions as the Base1 and
Short1 models; one simulation, L0.3, crashes before finishing due to
one sink particle growing to a mass greater than that of \sgra, so I
reran it with initial conditions identical to those of model
Base2. All simulations ran for 6 Myr, the presumed time interval since
the start of the AGN phase.

\section{Results}\label{sec:results}

I first present the results of the Control simulation, which has no
AGN activity and is designed to show the stability of the initial gas
configuration. Then I present the five Base models, which all have the
same AGN parameters: an Eddington-limited outburst lasting for $t_{\rm
  q} = 1$~Myr. I discuss the morphology and the structures forming in
the perturbed CMZ, derive fragmentation rates and properties of the
major sink particle clusters. Later, I comment on the effect of
varying the AGN outburst duration and luminosity.

\begin{figure}
  \centering
    \includegraphics[width=0.48 \textwidth]{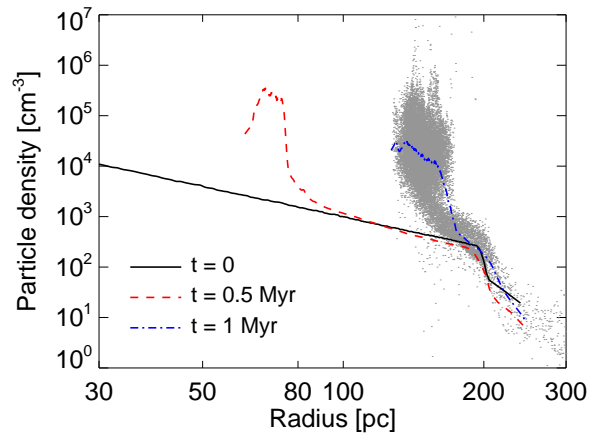}
  \caption{Radial density profiles of CMZ gas in the model Base1 at $t = 0$ (black), $t = 0.5$~Myr (red) and $t = 1$~Myr (blue). Grey points are a sample of SPH particle densities at $t = 1$~Myr.}
  \label{fig:base_radprof_early}
\end{figure}

\begin{figure*}
  \centering
    \includegraphics[trim = 4mm 23mm 8mm 0, clip, width=0.32 \textwidth]{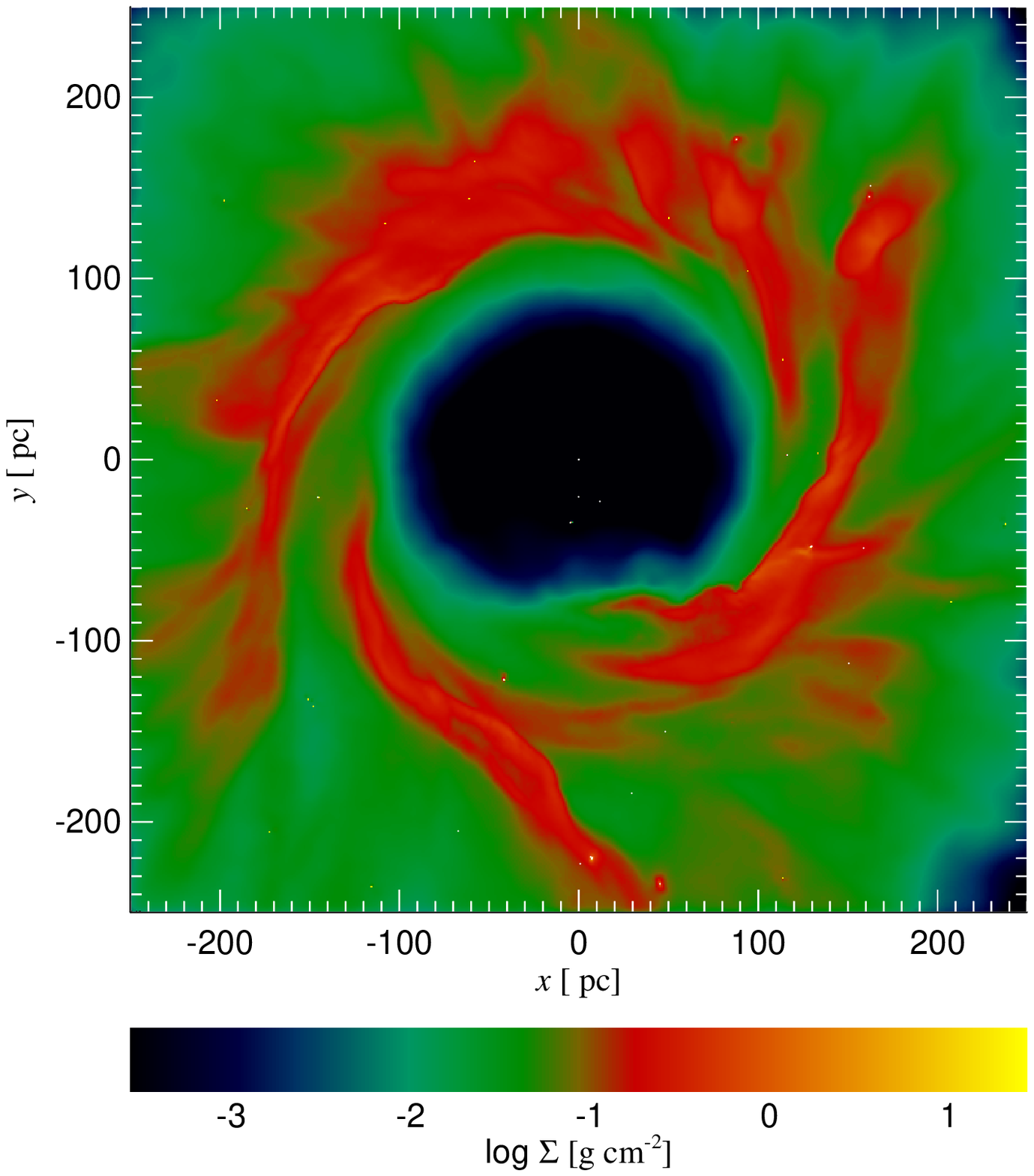}
    \includegraphics[trim = 4mm 23mm 8mm 0, clip, width=0.32 \textwidth]{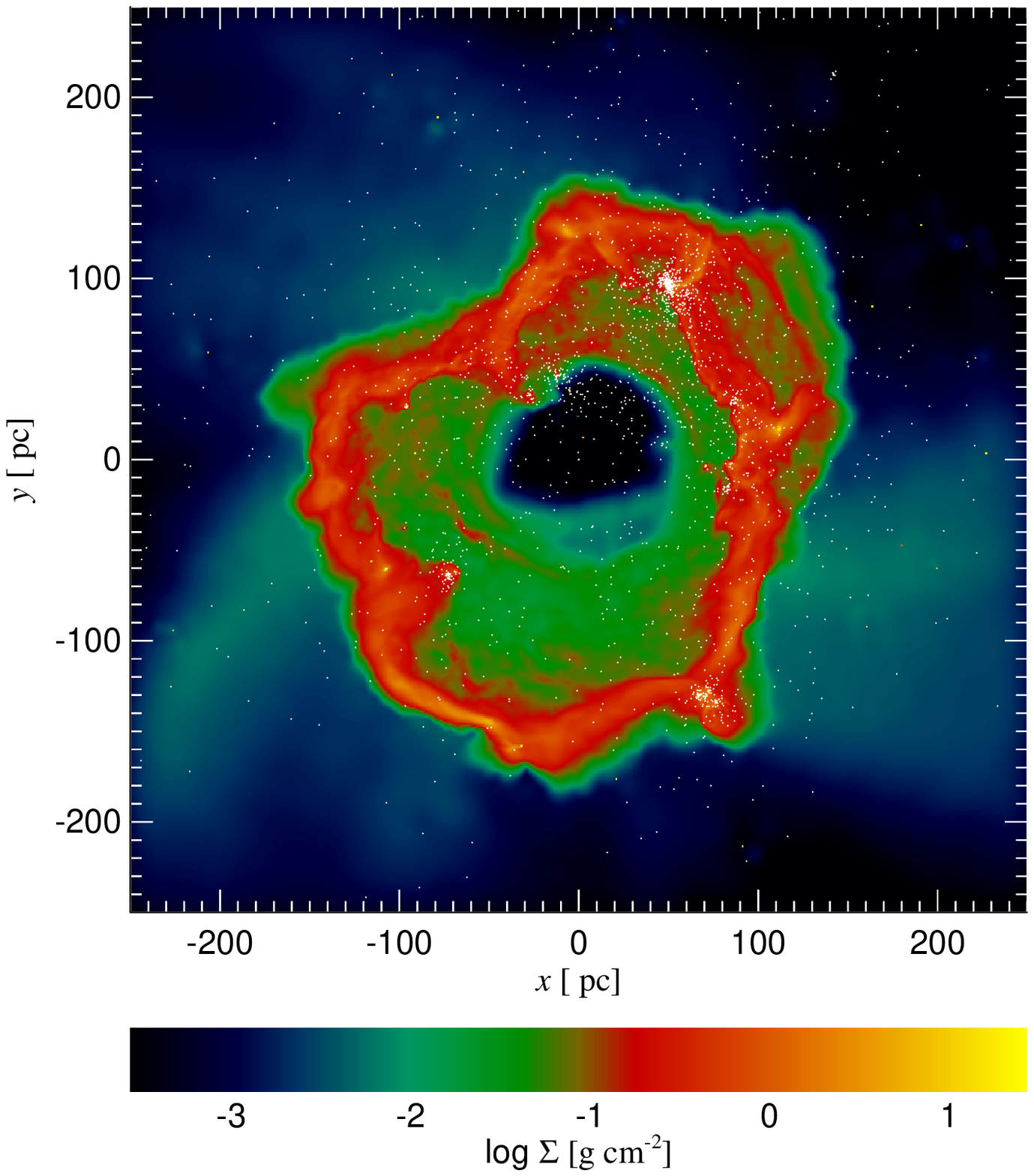}
    \includegraphics[trim = 4mm 23mm 8mm 0, clip, width=0.32 \textwidth]{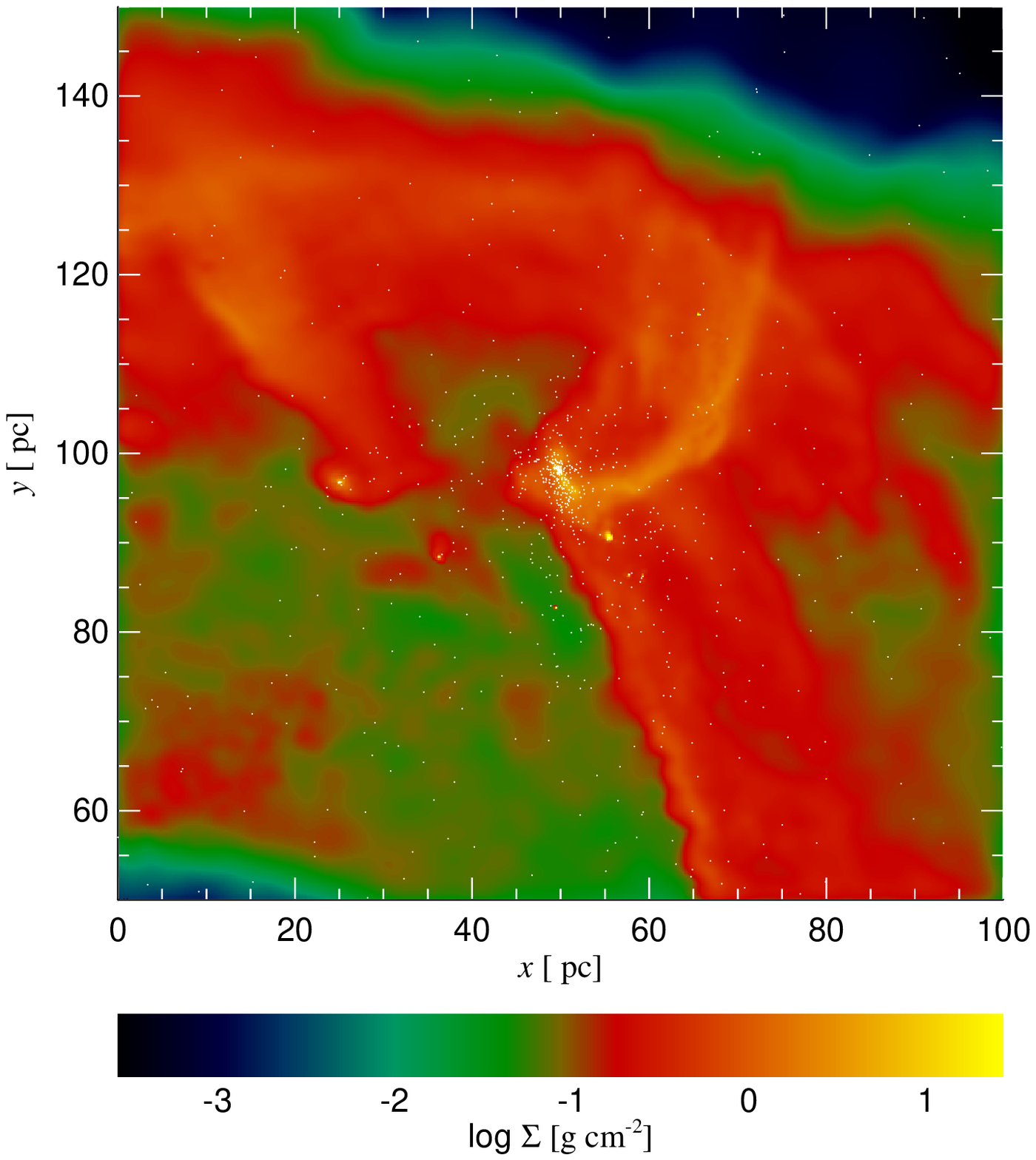}
  \caption{Projected gas density of fiducial run Base1 after the AGN
    switches off. All images show top view. {\bf Left}: $t = 3$~Myr;
    spiral perturbations are falling inward and facilitating radial
    migration of CMZ gas. {\bf Middle}: $t = 6$~Myr; a narrow
    asymmetric ring of gas is winding around the Galactic
    centre. White dots show the sink particles; several clusters can
    be identified. {\bf Right}: A zoom-in on the most massive sink
    particle cluster.}
  \label{fig:morph_late}
\end{figure*}

\subsection{Control simulation}

In this simulation, the gas disc is allowed to evolve freely without
external perturbation. Soon after the start of the simulation, a
circular overdensity forms in the centre of the ring and moves outward
with velocity $v \sim 40$~km/s; it reaches the outer edge of the disc
at $t \simeq 4$~Myr, at which point the disc radius has shrunk to
$\simeq 160$~pc. This happens because the initial conditions feature
sharp density cutoffs at both the inner and outer edge. As the disc
relaxes, density waves form, but the one on the outer disc is much
weaker due to the lower density contrast.

Subsequently, starting from $\sim 1$~Myr after the start of the
simulation, spiral density perturbations form in the inner part of the
disc. By $t=6$~Myr, they are present throughout the disc, with a
stronger density contrast in the inner $\simeq 80$~pc, reflecting the
shorter dynamical timescale there. The formation of spiral density
waves is expected, since the Toomre Q factor of the initial
configuration is $2.5$, around the limit where it becomes unstable to
such perturbations ($Q < 2$).

Figure \ref{fig:control_morph} shows the projected gas density map of
the disc at $t = 6$~Myr. Despite the presence of both circular and
spiral density waves, no fragmentation has occurred in the CMZ, and no
sink particles formed. This lack of fragmentation happens partially
due to the imposed temperature floor of $10^4$~K; if the gas could
cool down to much lower temperatures, some fragmentation would be
expected. However, stellar feedback and external radiation might keep
the gas in this configuration above the temperature required for
fragmentation. I discuss the importance of the gas equation of state
in Section \ref{sec:eos}.

\subsection{Fiducial runs: $l = 1$, $t_{\rm q} = 1$ Myr}

\begin{figure}
  \centering
    \includegraphics[width=0.48 \textwidth]{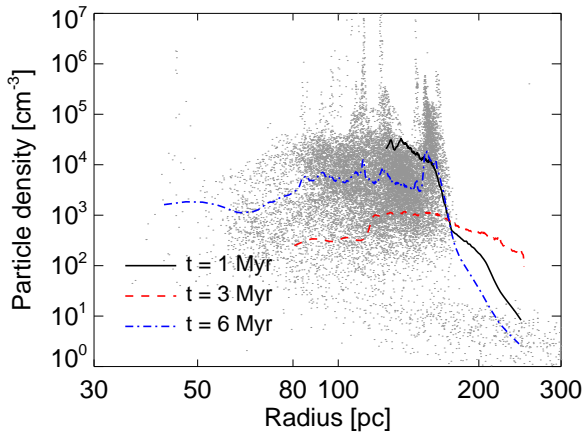}
  \caption{Radial density profiles of CMZ gas in the model Base1 at $t = 1$~Myr (black), $t = 3$~Myr (red) and $t = 6$~Myr (blue). Grey points are a sample of SPH particle densities at $t = 6$~Myr.}
  \label{fig:base_radprof_late}
\end{figure}

\subsubsection{Feedback-driven expansion}

From the start of the simulation, AGN wind impacts the inner edge of
the CMZ disc and pushes it outward. The expansion velocity is
initially approximately constant, $v_{\rm r,out} \simeq 150$~km/s $=
\sqrt{2}\sigma$. Formally, the purely energy-driven velocity of
expansion should be $\sim 330$~km/s, while purely momentum-driven
velocity should be $\sim 25$~km/s. The actual velocity is in between
the two, suggesting that a large fraction ($\sim 80\%$) of the energy
is reflected away from the dense gas and works to form bubbles in the
diffuse halo \citep[this effect was investigated in more detail
  by][]{Zubovas2014MNRASb}. Gas temperature at the edges of the CMZ
increases to $>10^5$~K, resulting in some gas evaporating from the
surfaces and escaping into the diffuse halo; meanwhile, most of the
gas stays close to the temperature floor. After $\sim 0.5$~Myr (Figure
\ref{fig:morph_early}, left panel) gas in the inner parts of the CMZ
is compacted so much that vertical gravitational collapse begins, as
the Toomre Q parameter in that region drops to $<1$. The inner parts
of the disc become thinner and less exposed to the AGN wind (red line
in Figure \ref{fig:base_radprof_early}); meanwhile, parts further away
become more exposed and start moving radially outward. Globally, this
results in a transition region forming between the evacuated inner
part of the disc and the stable outer parts. A surface layer of higher
density forms in the outer part of the disc, where the disc gas is
pushed inward by the expanding hot bubble in the halo (Figure
\ref{fig:morph_early}, middle panel).

By the end of the AGN outburst, i.e. at $t = 1$~Myr, the transition
region extends from $\sim 130$ to $\sim 160$~pc (Figure
\ref{fig:morph_early}, right panel; also blue line and grey points in
Figure \ref{fig:base_radprof_early}).  Column densities in the
transition region exceed $1$~g~cm$^{-2}$; this is enough to bring the
Toomre Q parameter below 1 ($\Sigma_{\rm crit} \simeq 0.4
R_{100}^{-1}$~g~cm$^{-2} \simeq 1700 R_{100}^{-1} \msun$~pc$^{-2}$).
As a result, spiral perturbations and self-gravitating clumps form
within the transition region.

All five Base models evolve very similarly during this stage, since
their evolution is dominated by a global process (AGN feedback),
rather than local gravity.

\begin{figure}
  \centering
    \includegraphics[width=0.48 \textwidth]{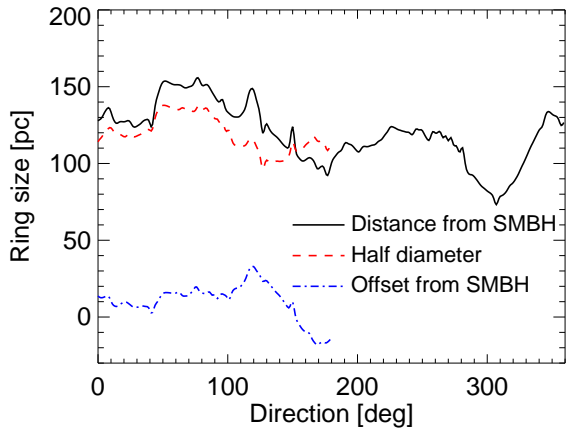}
  \caption{Average radius (black solid line), half-diameter (red
    dashed line) and displacement (blue dot-dashed line) of the dense
    gas ring, defined as $n > 100$~cm$^{-3}$, as function of direction
    counterclockwise from left in model Base1 at $t = 6$~Myr. Gas
    further than $200$~pc from the centre was excluded when
    calculating these values. The radius (i.e. distance from the
    origin to the dense ring) varies between $\sim 80$ and $\sim
    150$~pc. The half-diameter (half the distance
    between the ring on opposite sides of the origin) varies between
    $\sim 100$ and $\sim 140$~pc, and the SMBH is displaced from the
    centre of the ring by about $20$~pc.}
  \label{fig:base1_morph}
\end{figure}

\subsubsection{Fragmentation}

Once AGN activity ends, the CMZ material collapses back toward the
centre. The circularization radius is $\sim 60$~pc, but falling
material overshoots this limit and moves to radial distances of $\sim
40$~pc before expanding again. In effect, individual gas particles
move on eccentric orbits and the whole disc appears to breathe. The
particles on the outer edge of the transition region at $t = 1$~Myr
continue moving outward for a while, encompassing the whole disc and
causing it to expand to a maximum radius of $\sim 270$~pc at $t = 2$
Myr.

During this time, spiral perturbations continue to grow in the
disc. By $t = 2$~Myr, they are clearly visible. By $t = 3$~Myr, they
have become significant density enhancements, facilitating infall of
material to small radii (Figure \ref{fig:morph_late}, left panel). The
formation and growth of perturbations is a stochastic process, and as
a result, differences among the five Base models begin to manifest. In
the rest of this subsection, the descriptions refer to model Base1,
unless otherwise noted.

By the end of the simulation, the CMZ gas is distributed strongly
anisotropically, with a narrow ($d_{\rm ring} \sim 20-40$~pc) ring
winding around the Galactic centre at distances ranging from $r_{\rm
  min} \sim 80$~pc to $\sim 150$~pc (top-left and bottom in the middle
panel of Figure \ref{fig:morph_late}, respectively; see also Figures
\ref{fig:base_radprof_late} and \ref{fig:base1_morph}). The ring has a
sharp outer edge, with density dropping by several orders of magnitude
outside $\sim 150$~pc, and a less pronounced inner edge where density
decreases by approximately a factor $4$. The ring can be roughly
approximated as an ellipse with semimajor axis of $140$~pc and
semiminor axis of $100$~pc, with the semimajor axis making a $30^o$
angle with the Y axis. The SMBH is displaced by $\sim 20$~pc from the
geometric centre of the ellipse. The total mass of the gas ring,
defined as having gas with density $n_{\rm dense} > 100$~cm$^{-3}$, is
$M_{\rm dense} = 6.9 \times 10^7 \; \msun$. This value does not
include the mass of sink particles (see below). The cutoff density is
chosen in such a way that if the selected gas cools down isobarically
to $20$~K, the column density through the ring would be $N_{\rm dense}
\simeq n_{\rm dense} \times \left(T_{\rm floor}/ 20 {\rm
  K}\right)^{2/3} \times d_{\rm ring} \simeq 4 \times
10^{23}$~cm$^{-2}$, equal to the threshold used when identifying the
ring from infrared observations \citep{Molinari2011ApJ}.

\begin{figure}
  \centering
    \includegraphics[trim = 4mm 23mm 8mm 0, clip, width=0.48 \textwidth]{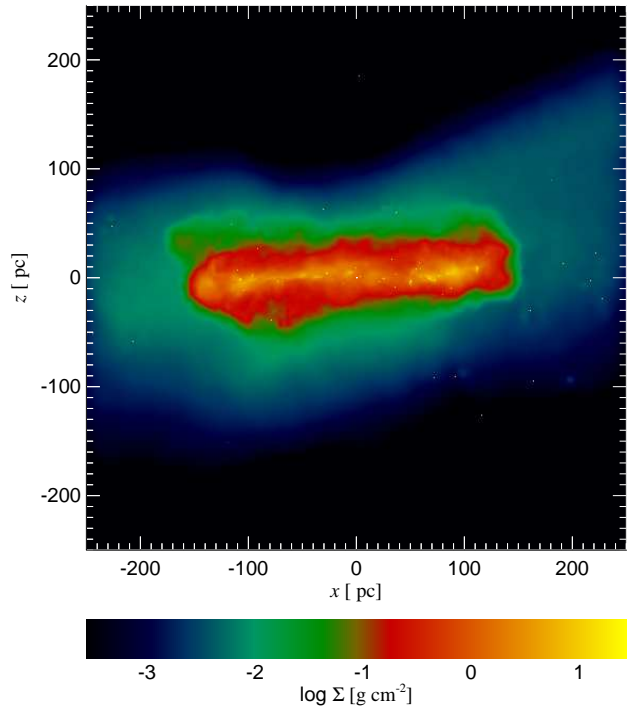}
  \caption{Side view of model Base1 at $t = 6$~Myr. The gas ring is
    slightly disturbed, but generally stays on the $Z=0$ plane.}
  \label{fig:base_side}
\end{figure}

\begin{figure*}
  \centering
    \includegraphics[trim = 4mm 23mm 8mm 0, clip, width=0.24 \textwidth]{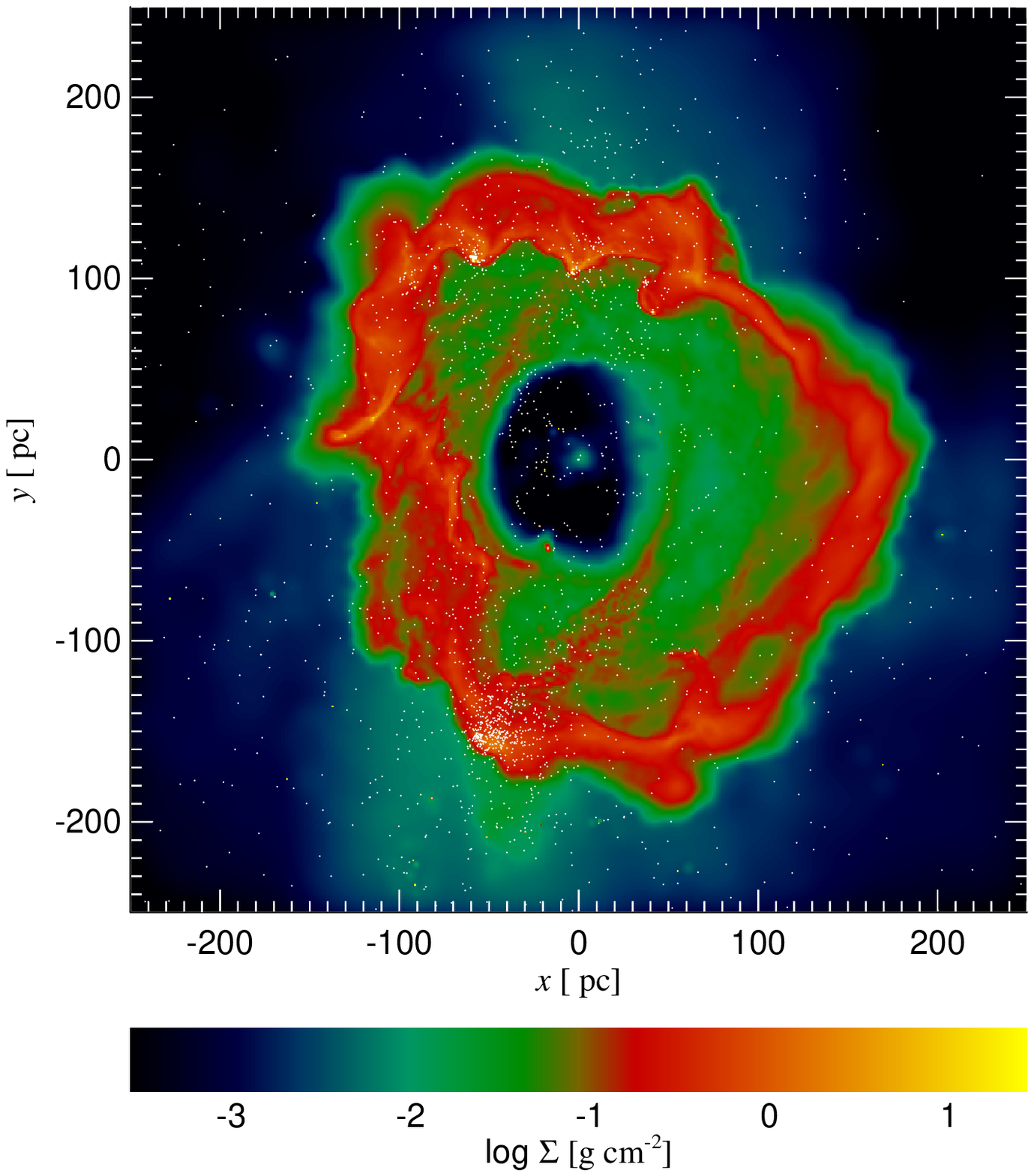}
    \includegraphics[trim = 4mm 23mm 8mm 0, clip, width=0.24 \textwidth]{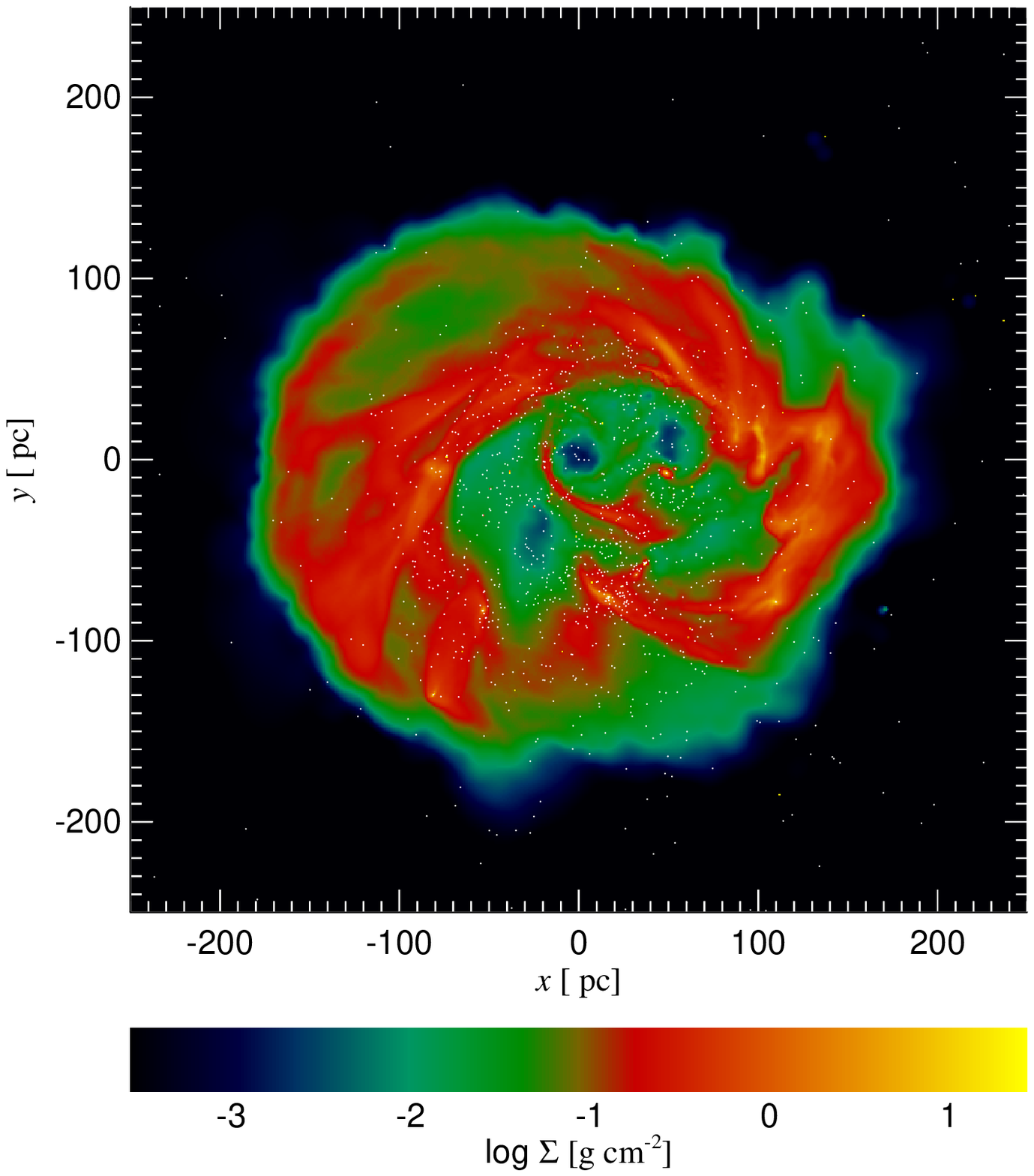}
    \includegraphics[trim = 4mm 23mm 8mm 0, clip, width=0.24 \textwidth]{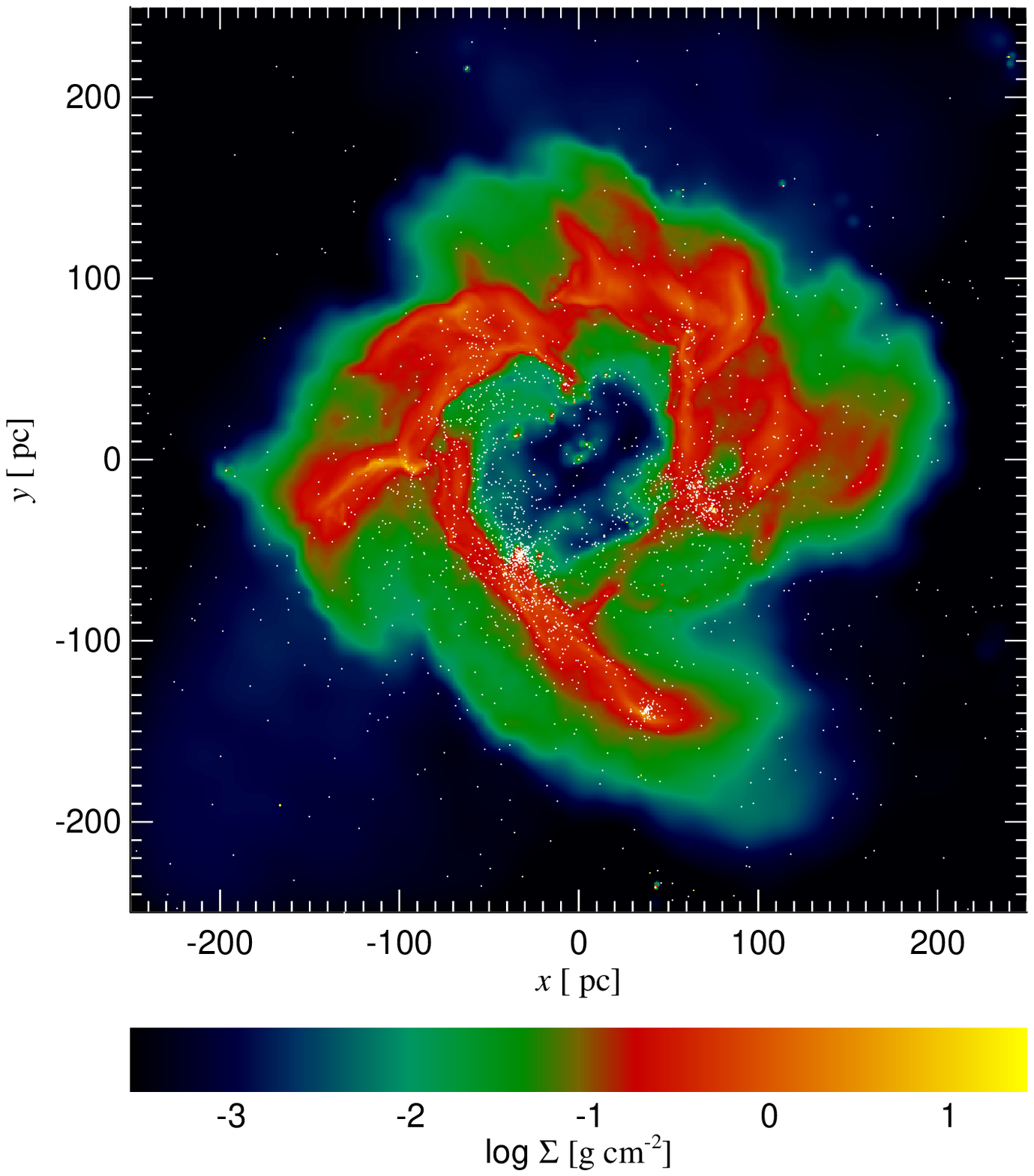}
    \includegraphics[trim = 4mm 23mm 8mm 0, clip, width=0.24 \textwidth]{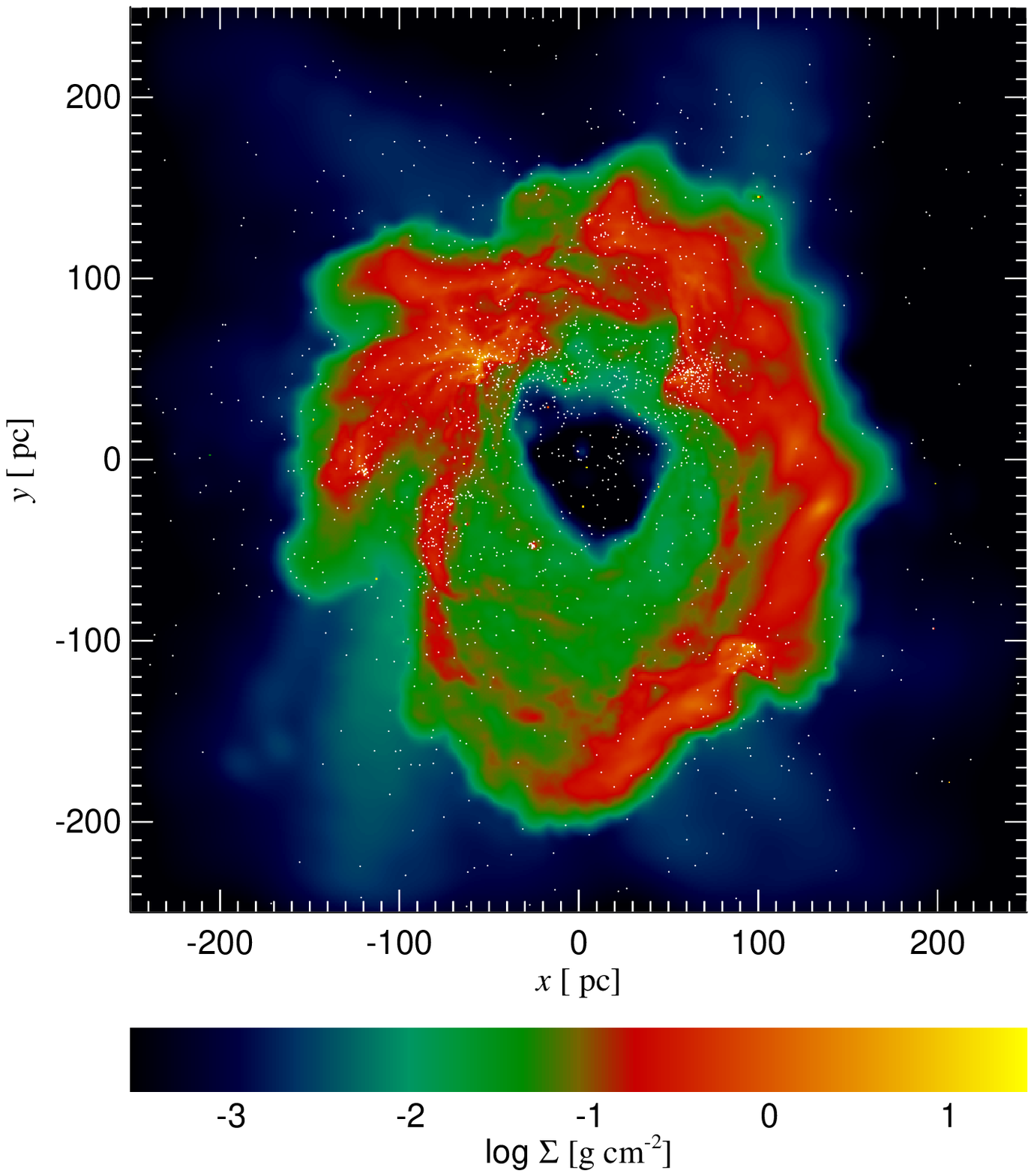}
  \caption{Projected gas column density of models Base2 (left panel), Base3
    (second panel), Base4 (third panel) and Base5 (last panel) at $t =
    6$~Myr. The gas morphologies differ significantly, with no
    well-defined rings in any of the models.}
  \label{fig:base_var}
\end{figure*}

\begin{figure}
  \centering
    \includegraphics[width=0.48 \textwidth]{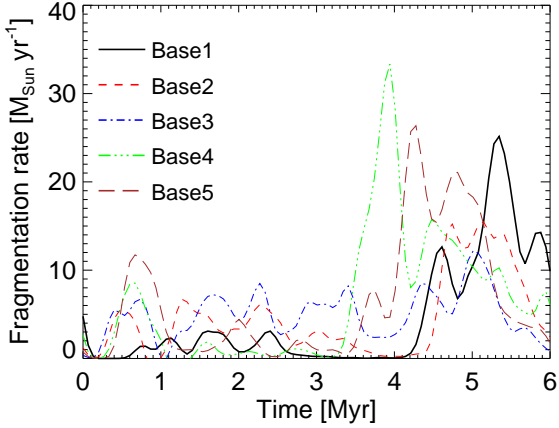}
  \caption{Fragmentation rates in the five Base models. The algorithm
    calculates the change in total sink particle mass, so the actual
    rates may be slightly higher if many sink particles are accreted
    by the central SMBH. All models except Base3 show qualitatively
    similar behaviour - initially small fragmentation rates lead to a
    surge once the CMZ ring begins fragmenting in earnest, at
    $3.5-4.5$~Myr. In the Base3 model, fragmentation rates stay roughly
    the same throughout the simulation.}
  \label{fig:base_fragrate}
\end{figure}

The densest parts of the perturbations become self-gravitating clumps
and create sink particles. The first sink particles appear before the
AGN switches off, but fragmentation starts in earnest at $t \ga
4.5$~Myr. There are five regions of intense fragmentation, lighting up
one after another. In addition, some sink particles form in the dense
filaments; these most likely represent small star clusters and
associations with masses at or below the simulation resolution of
$8000 \msun$. By the end of the simulation, the total mass in sink
particles is $M_{\rm sink} = 2.8 \times 10^7 \msun$. Thus the total
mass in sink particles and dense gas is $\sim 9.7 \times 10^7 \msun$,
i.e. almost all of the initial CMZ material ends up either converted
into sink particles or in the dense gas ring. The presence of dense
gas is not unexpected, since the initial conditions contain gas denser
than the threshold used to identify the ring; however, the
transformation from a smooth disc into a narrow ring is
remarkable. The average fragmentation rate over the $1.5$~Myr period
of intense fragmentation is $\dot{M}_{\rm frag} \simeq 15 \;
\msun$~yr$^{-1}$, but it can rise up to $\sim 25 \; \msun$~yr$^{-1}$
(see Figure \ref{fig:base_fragrate}). Assuming a typical star
formation efficiency on such mass scales of $2-10\%$ gives a star
formation rate $\dot{M}_* \simeq 0.25 - 1.25 \; \msun$~yr$^{-1}$,
dropping below $0.2 \; \msun$~yr$^{-1}$ at $t = t_{\rm n} = 6$~Myr.

Most sink particles move far away from the regions where they form;
this is most likely a numerical effect arising due to the timesteps
and smoothing lengths of the sink particles being too large for proper
orbit integration within the clusters. However, one can still identify
five or six clusters (Figure \ref{fig:morph_late}, middle panel),
distributed asymmetrically within the CMZ. The most massive such
cluster is located at $\lbrace X,Y\rbrace \simeq \lbrace 50,
100\rbrace$~pc (visual estimate; Figure \ref{fig:morph_late}, right
panel). By measuring the cumulative mass within a given distance from
the centre of the cluster, I estimate its size to be $R_{\rm cl} \sim
10$~pc and mass $2.8 \times 10^6 \; \msun$. A more rigorous estimate
of cluster size is not reasonable due to lack of numerical resolution
and appropriate star formation physics within the model. Other
clusters have sizes $\sim 5$~pc or less and masses $\sim 0.7-1.6
\times 10^6 \; \msun$. The typical star formation efficiency leads to
star cluster mass estimates of $1.4 \times 10^4 - 2.8\times 10^5 \;
\msun$, very similar to the masses of the observed clusters. The
orbits of the sink particle clusters, much like the gas orbits, are
non-circular and may be strongly irregular. This is consistent with
observations of the motion of both Arches \citep{StolteEtal08} and
Quintuplet \citep{Stolte2014ApJ} clusters.

The gas and sink particles are distributed almost symmetrically around
the $Z=0$ plane. In Figure \ref{fig:base_side}, I show the gas column
density in a side view of the CMZ. Small variations in gas height and
position are visible across the CMZ, but the displacement from $Z=0$
does not exceed the ring thickness $h \la 40$~pc.

The other four Base models exhibit very different gas morphologies in
the XY plane. Figure \ref{fig:base_var} shows the gas distribution in
those models at $t = 6$~Myr. All four gas distributions are clumpy,
but the large-scale gas distribution differs. Model Base2 (left panel)
has a full ring, model Base5 (right panel) shows approximately two
thirds of a ring with a large gap in the bottom-left direction, model
Base4 has prominent spiral streamers which may coalesce into a ring
structure in several Myr. Finally, model Base3 has a very weak (low
density contrast) and highly off-centre elliptical structure, with
very dense gas at the apocentre in the positive-X direction. It is
interesting to note that even though the morphologies of the five
models differ significantly, the formal ratio of semimajor to
semiminor axis lengths, when defined as the maximum ratio of mean
diameters of the dense gas features in two perpendicular directions,
is rather similar, ranging from 1.34 to 1.53 (Table \ref{table:param},
last column).

The morphological differences are reflected in the fragmentation rates
(Figure \ref{fig:base_fragrate} and Table \ref{table:param}, columns
4-6). In models Base4 and Base5, several prominent bursts of
fragmentation occur earlier than in Base1, starting at $t \simeq
3.4-4.0$~Myr. The peak fragmentation rates in these models are
slightly higher than those in Base1, while the mean rates over the
significant fragmentation period are similar. In model Base2,
significant fragmentation starts at a similar time as in Base1, but
proceeds slower. In model Base3 the fragmentation rate stays
approximately constant, between $5-10 \; \msun$~yr$^{-1}$, throughout
the simulation. The total mass of sink particles varies from
$2.7\times 10^7 \msun$ in Base1 and Base2 to $3.8\times 10^7 \msun$ in
Base4 and Base5, with an average value $(3.2 \pm 0.5) \times 10^7
\msun$. The vertical structure of all four models is very similar to
that of the Base1.

\subsubsection{Mixing of gas and angular momentum} \label{sec:mixing}

\begin{figure}
  \centering
    \includegraphics[width=0.48 \textwidth]{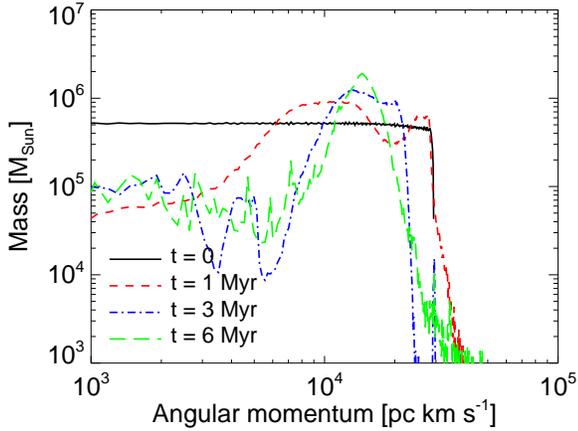}
  \caption{Distribution of CMZ gas angular momentum in the model Base1
    at $t = 0$, $1.5$, $3$ and $6$~Myr (black solid, red dashed, blue
    dot-dashed and green long-dashed lines, respectively). The gas
    initially has a flat angular momentum distribution, which becomes
    progressively more squashed toward a single peak at $l \sim 1.5
    \times 10^4$~pc~km/s.}
  \label{fig:angmom}
\end{figure}

A general outcome of the five Base simulations is that an initially
smooth and stable distribution of gas becomes clumpy and
anisotropic. This can be understood as a consequence of mixing of gas
and its angular momentum. During the AGN phase, a large amount
of gas (almost the whole CMZ) is pushed together into a narrow
ring. This both increases the surface density of this gas, making it
prone to gravitational instabilities, and efficiently mixes the
angular momentum of gas, ensuring that the ring disperses slowly
after the AGN switches off.

To show this, I plot the angular momentum distribution of gas at $t =
0$, $t = 1$~Myr (end of AGN phase), $t = 3$~Myr and $t = 6$~Myr in
Figure \ref{fig:angmom}. Initially, the angular momentum distribution
is flat throughout the disc, since each annulus contains the same mass
of material. AGN feedback pushes gas with low angular momentum
(central parts of the disc) to merge with the high angular momentum
gas, leading to a bump at middle values $l \sim 10^4$~pc~km/s; the
outer edge of the disc is not perturbed radially at $t =
1$~Myr. Later, this process continues, and by $t = 3$~Myr, the bump
has moved to larger values due to mixing with the material at the
outskirts of the disc. Even by the end of the simulation, material is
still being pushed together by spiral perturbations and gravitational
instabilities; there is, however, a growing high angular momentum
tail, which arises due to viscous spreading of the ring material. The
total mass of material with very low angular momentum increases after
the AGN switches off; this represents clumps of gas falling inward and
coming very close to the SMBH, potentially feeding it (see Section
\ref{sec:fuelling}).

\subsection{Shorter AGN episode models: $t_{\rm q} = 0.7$ Myr} \label{sec:short}

The five Base models consistently produce gas rings, or parts of
rings, that are too large and form sink particle clusters too late as
compared with observations of the CMZ structure. Motivated by this, I
run four models with $t_{\rm q} = 0.7$~Myr, labelled Short1 to Short4
(see Table \ref{table:param} for details and main results). As before,
the models have stochastically different initial conditions, leading
to different fragmentation histories and gas morphologies at the end
of the simulation.

\begin{figure*}
  \centering
    \includegraphics[trim = 4mm 23mm 8mm 0, clip, width=0.24 \textwidth]{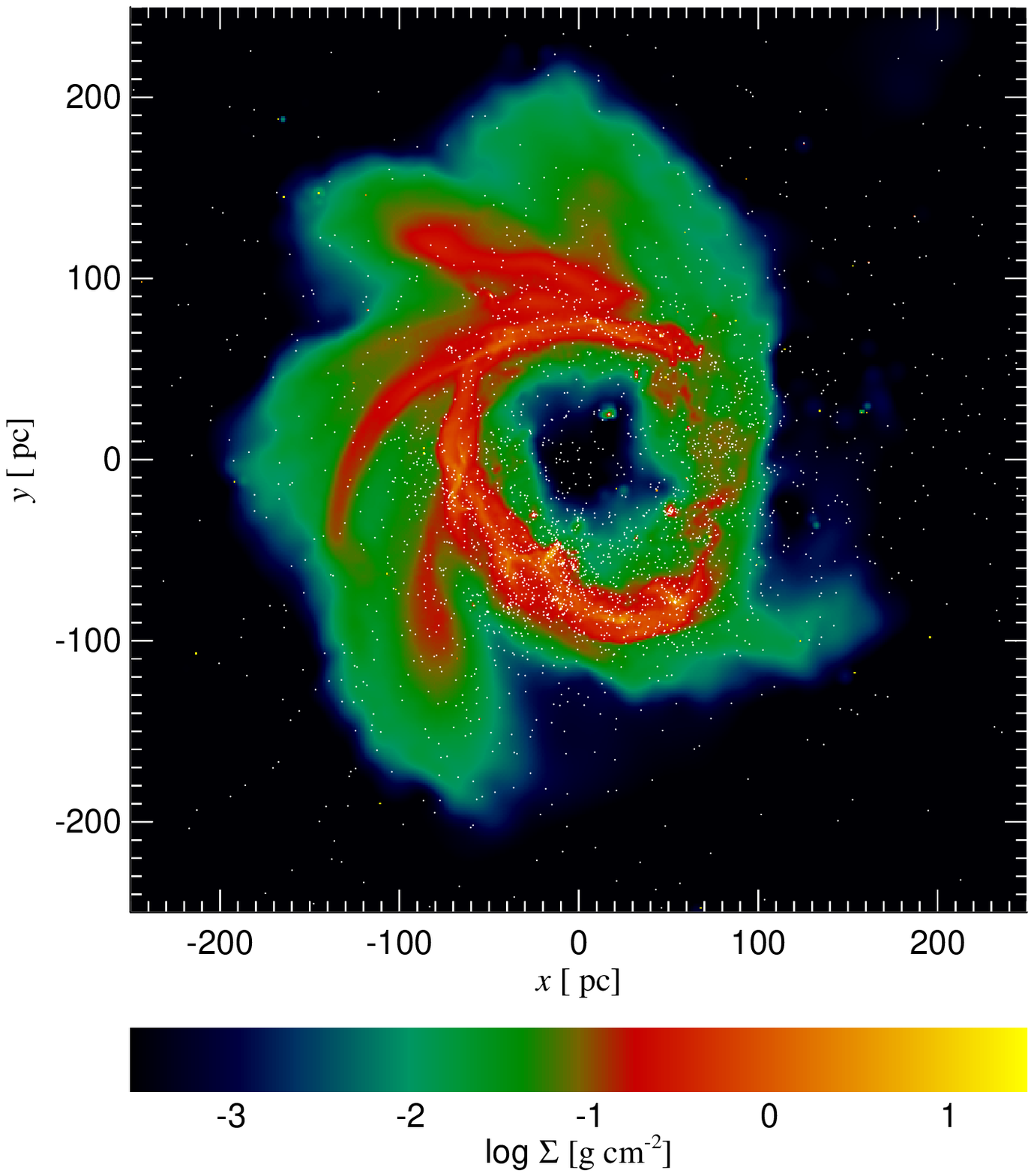}
    \includegraphics[trim = 4mm 23mm 8mm 0, clip, width=0.24 \textwidth]{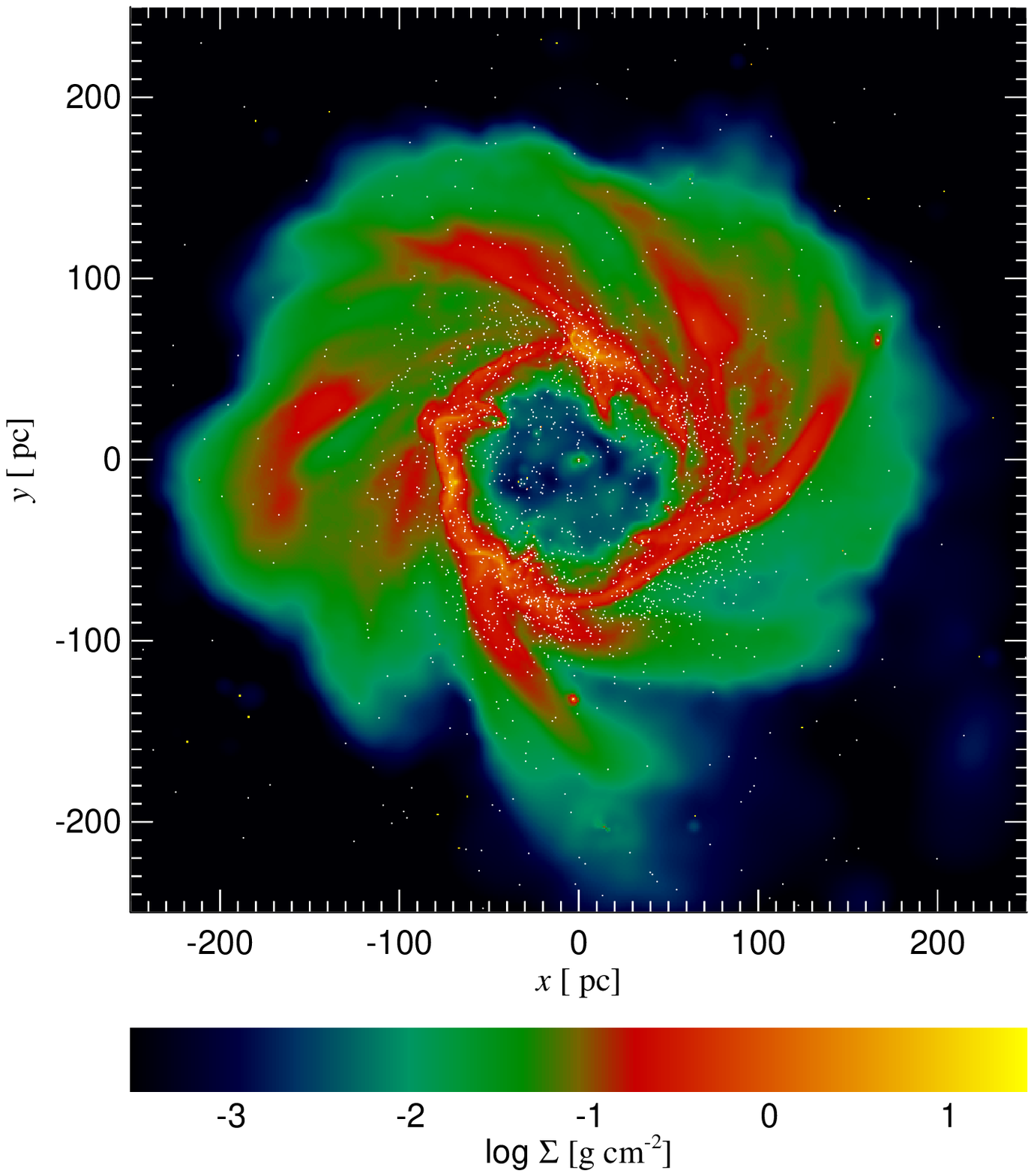}
    \includegraphics[trim = 4mm 23mm 8mm 0, clip, width=0.24 \textwidth]{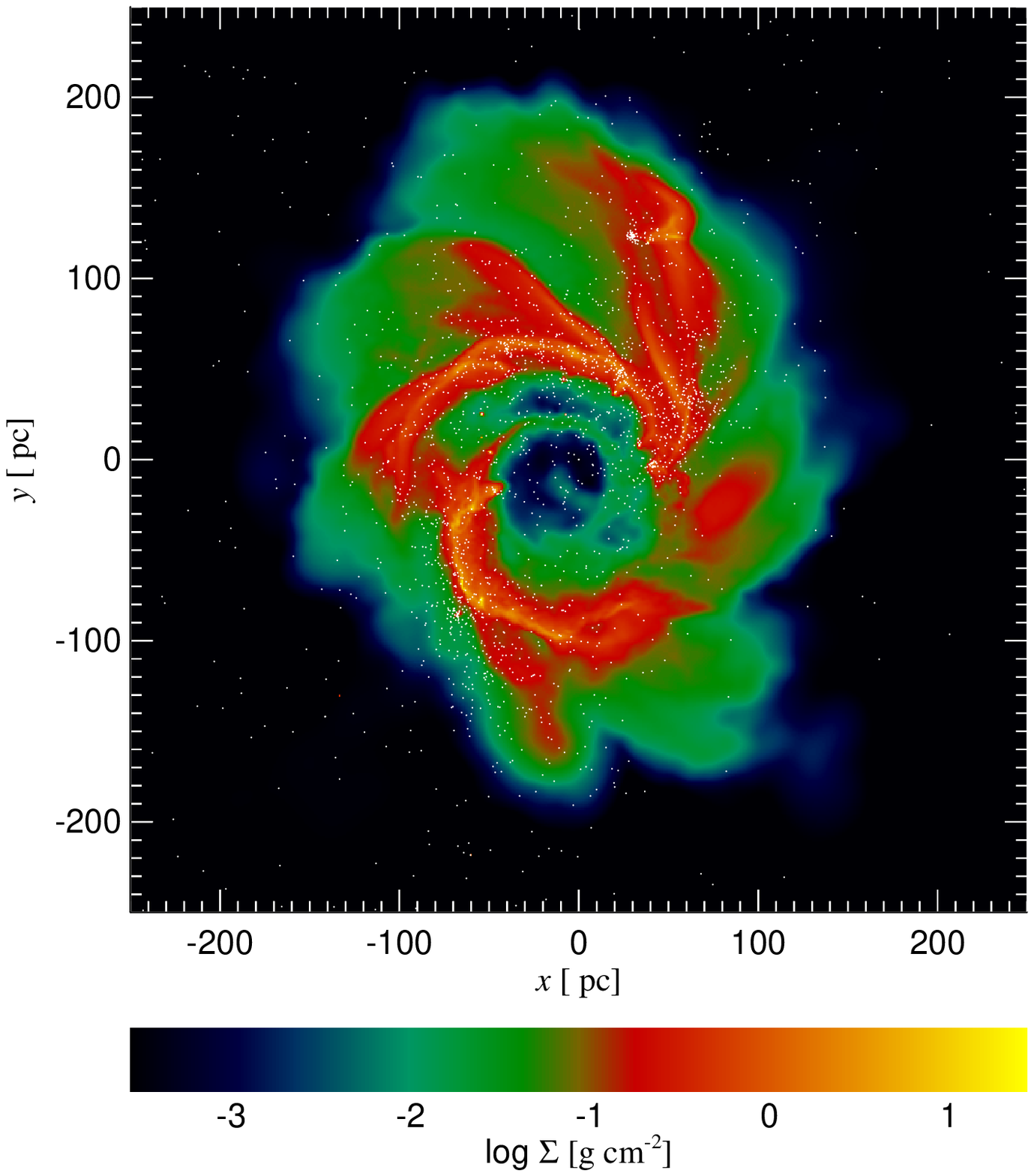}
    \includegraphics[trim = 4mm 23mm 8mm 0, clip, width=0.24 \textwidth]{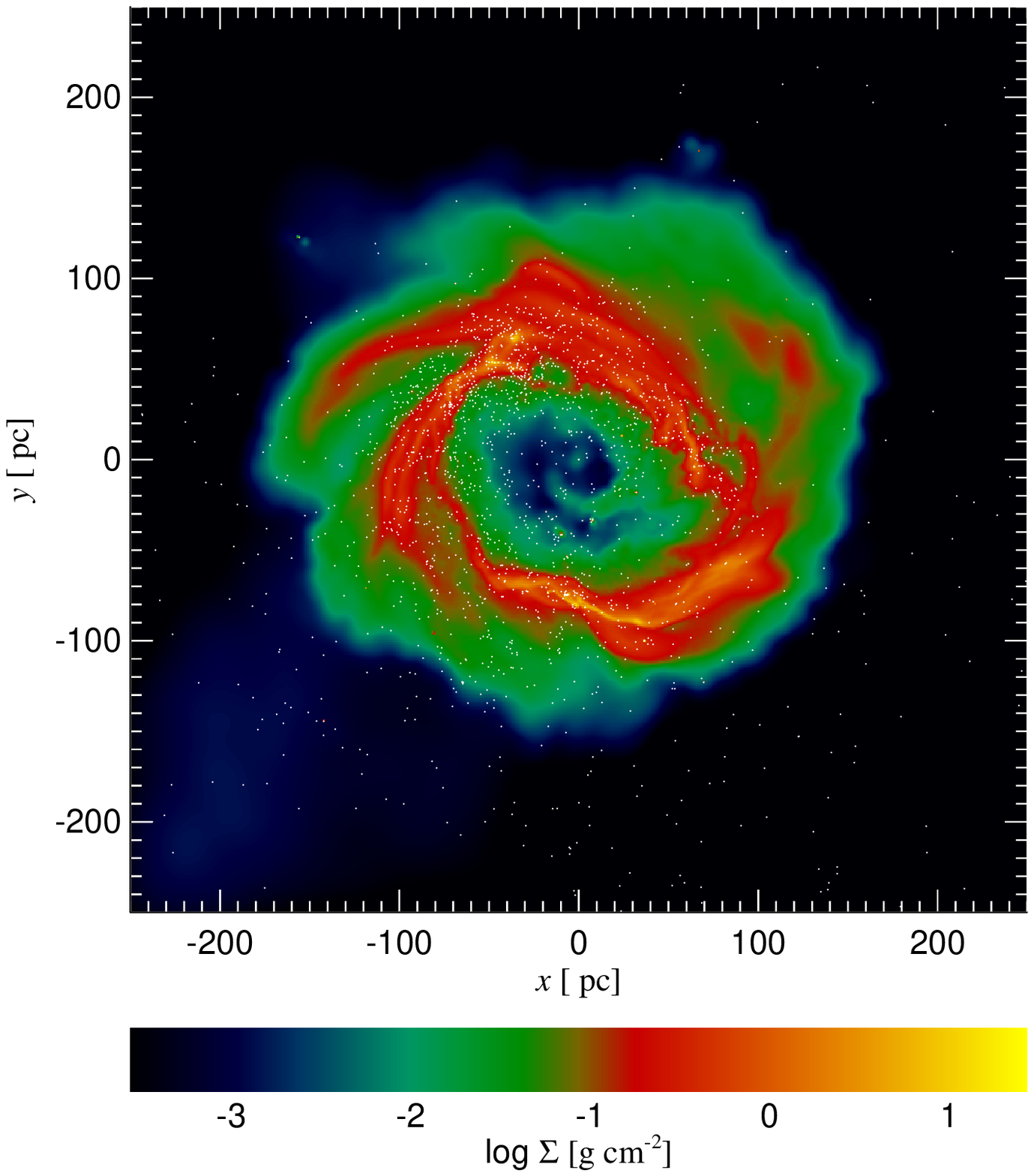}
  \caption{Projected gas column density of models Short1 (left panel), Short2
    (second panel), Short3 (third panel) and Short4 (last panel) at $t
    = 6$~Myr. The perturbed gas rings are smaller than in the Base
    models, with a variety of structures present.}
  \label{fig:short_morph}
\end{figure*}

\begin{figure}
  \centering
    \includegraphics[width=0.48 \textwidth]{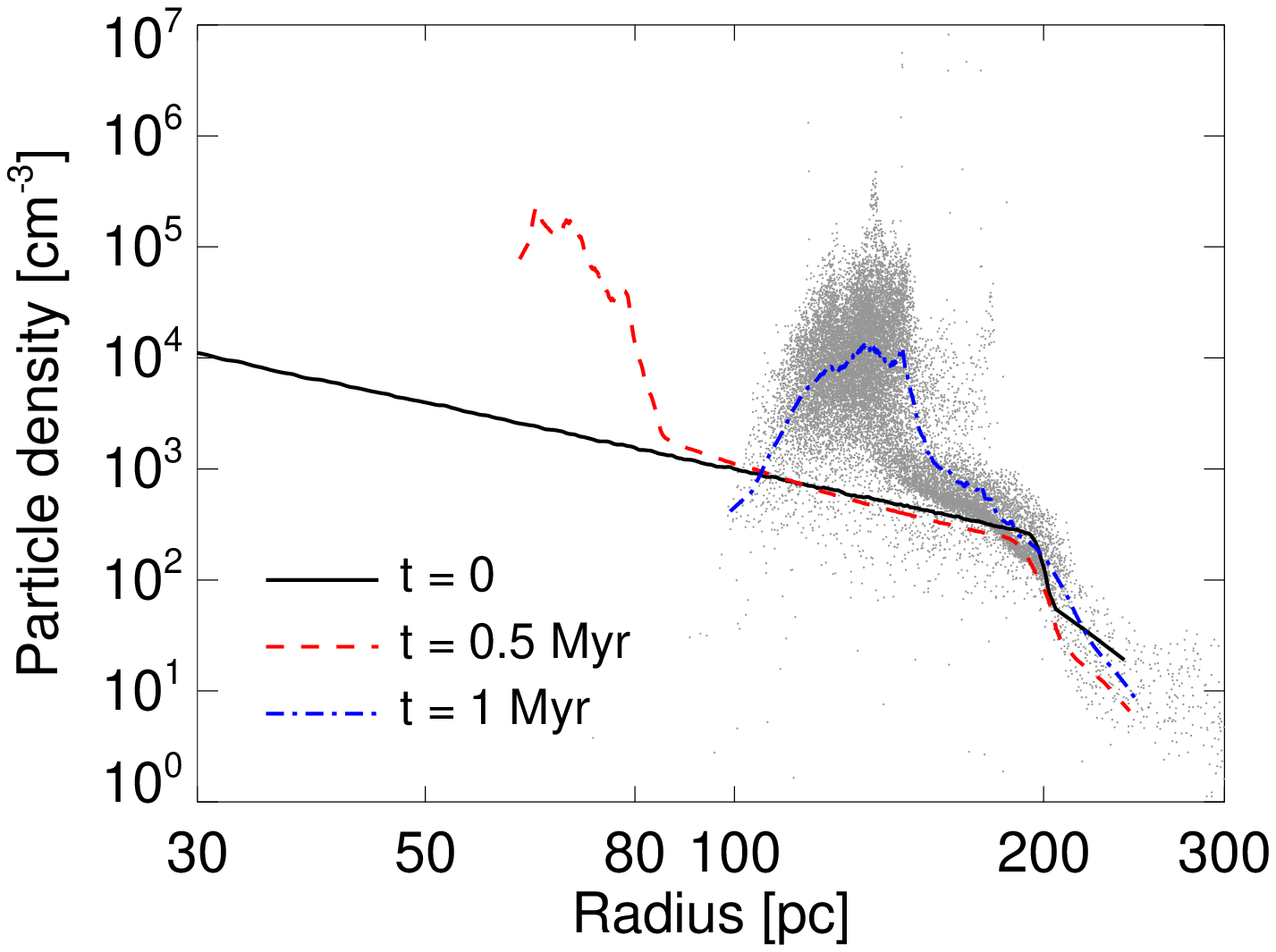}
    \includegraphics[width=0.48 \textwidth]{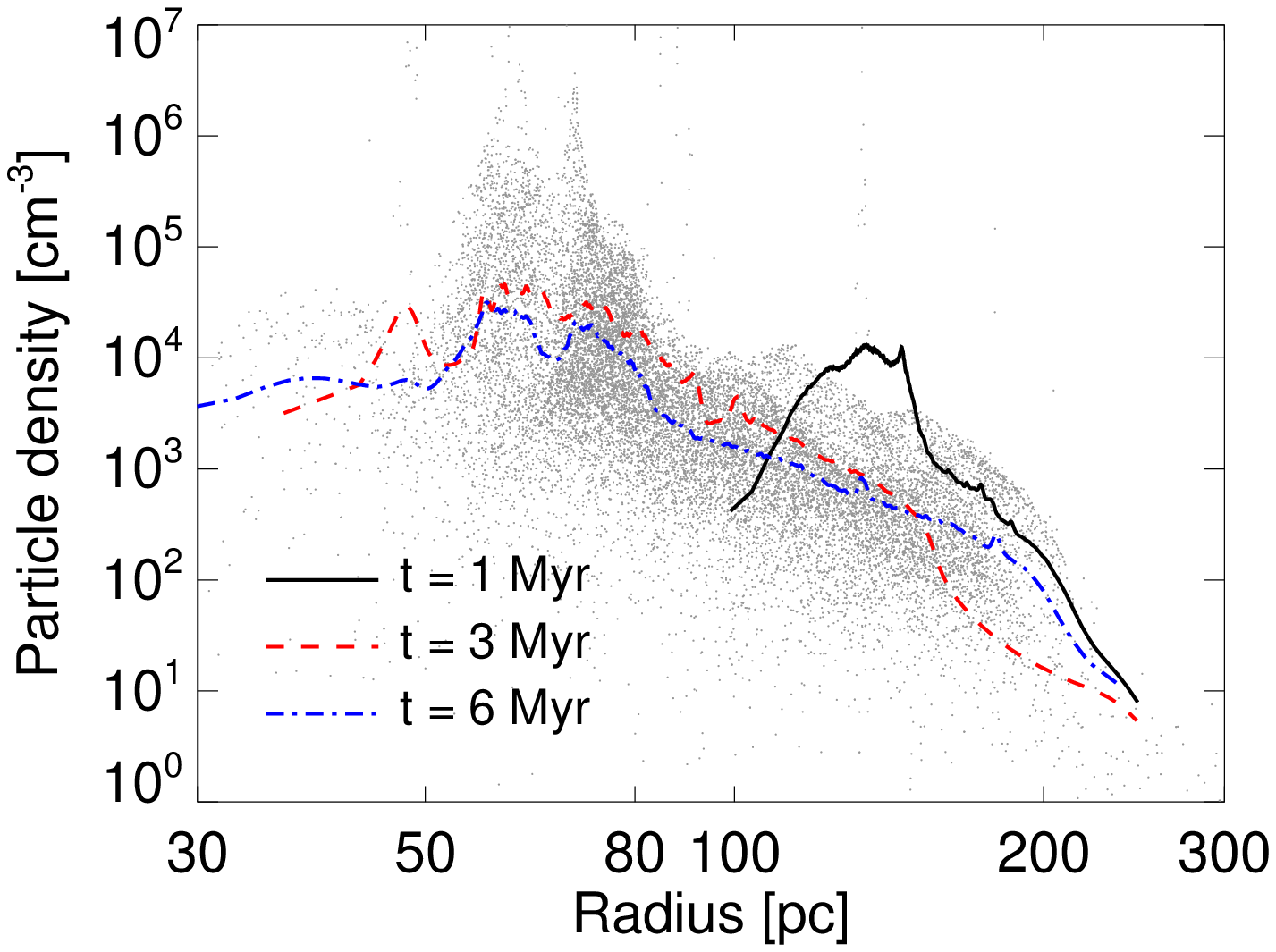}
  \caption{Radial density profiles of CMZ gas in model Short1. {\bf Top panel:} $t = 0$ (black solid), $t = 0.5$~Myr (red dashed) and $t = 1$~Myr (blue dot-dashed). Grey points are a sample of SPH particle densities at $t = 1$~Myr. {\bf Bottom panel:} $t = 1$~Myr (black solid), $t = 3$~Myr (red dashed) and $t = 6$~Myr (blue dot-dashed). Grey points are a sample of SPH particle densities at $t = 6$~Myr.}
  \label{fig:short_radprof}
\end{figure}

\begin{figure}
  \centering
    \includegraphics[width=0.48 \textwidth]{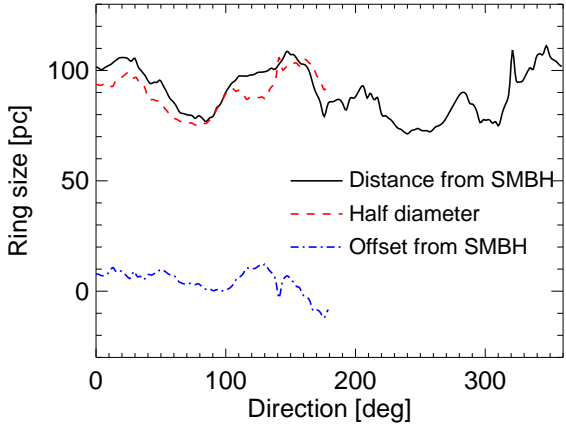}
  \caption{Average radius (black solid line), half-diameter (red
    dashed line) and displacement (blue dot-dashed line) of the dense
    gas ring, defined as $n > 100$~cm$^{-3}$, as function of direction
    counterclockwise from left in model Short4 at $t = 6$~Myr. The
    radius varies between $\sim70$ and $\sim110$~pc. The
    half-diameter varies between $\sim75$ and $\sim105$~pc, and the
    SMBH is displaced from the centre of the ring by about $10$~pc.}
  \label{fig:short_size}
\end{figure}

\begin{figure}
  \centering
    \includegraphics[width=0.48 \textwidth]{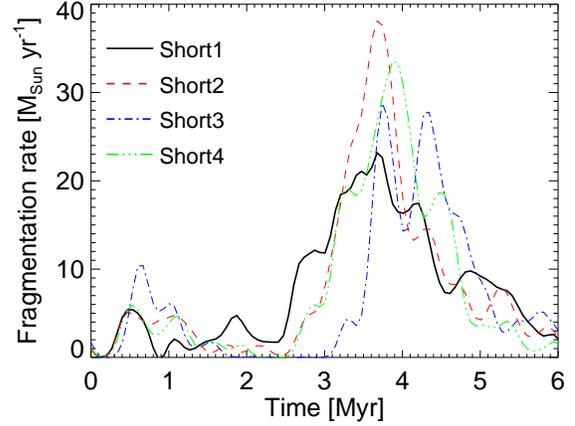}
  \caption{Gas fragmentation rate as function of time in the four
    Short models. The bursts of significant fragmentation occur at $t
    = 2.5-3$~Myr, earlier than in the Base models.}
  \label{fig:short_fragrate}
\end{figure}

Qualitatively, the evolution of the Short models is very similar to
that of the Base models. Initially, the AGN wind pushes disc gas
together into a narrow ring, which becomes gravitationally unstable
and fragments over several dynamical times. The shorter AGN episode
duration results in a smaller initial extent of the ring, which is
mirrored by the final sizes as well (see Figures
\ref{fig:short_morph}, \ref{fig:short_radprof} and
\ref{fig:short_size}). The ring eventually has a radial extent from
$\sim 50$ to $\sim 80$~pc from the centre, comparable to the observed
CMZ size. The density contrast on each edge is not as sharp as in the
Base models, but still noticeable, so that the ring can be identified
as such (Figure \ref{fig:short_radprof}, blue dashed line and points
in the bottom panel). The ring evolves faster: since the expansion of
the ring happens at constant velocity, and the subsequent evolution
depends on the dynamical time, which increases linearly with radius,
one would expect $t_{\rm frag, short}/t_{\rm frag,base} = t_{\rm q,
  short} / t_{\rm q,base} = 0.7$. In fact, rapid fragmentation begins
$2.5-3$~Myr after the start of the simulation (Figure
\ref{fig:short_fragrate}), almost $30\%$ earlier than in the Base
models, validating this simple explanation. The star clusters forming
during this period would have ages consistent with observations of the
Arches and Quintuplet clusters. Significant fragmentation has finished
in all models by $6$~Myr, so the expected star formation rate today is
$< 0.2 \msun$~yr$^{-1}$. The masses and sizes of the star clusters are
more difficult to determine than in the Base models, since the older
sink particle clusters have lost more members due to numerical
effects. The total masses of sink particles forming in the four
simulations are $2.4-2.9\times 10^7 \msun$, only slightly lower than
in the Base models.

The final shape of the gas rings varies among the four Short models
(Figure \ref{fig:short_morph}), but the rings are generally more
regular than in the Base simulations. Their ellipticity is similar to
the longer-duration runs, $a/b = 1.27-1.55$ (Table \ref{table:param},
last column), but the displacement of the SMBH from the centre
generally does not exceed 10 pc. As an example, I plot the radius,
half-diameter and displacement of the gas ring in the Short4 model in
Figure \ref{fig:short_size}. The ring half-diameter varies between
$75$ and $105$~pc, quite similar to the $60$ and $100$~pc of the model
by \citet{Molinari2011ApJ}. \sgra is displaced from the centre of the
ring by $\sim 10$~pc, less than in the \citet{Molinari2011ApJ} model,
but that value is less strongly constrained than the size of the ring
itself.

Overall, the Short models give a final gas distribution in better
agreement with the observed shape and size of the CMZ than the Base
models. However, the stochastic evolution of the gas ring results in
large uncertainties regarding various more detailed parameters (such
as the number, size and mass of star forming regions and young
clusters), preventing a more rigorous constraint on past AGN activity
from being made. I now turn to the four models designed to show the
effect of varying the AGN duration and luminosity more significantly.

\subsection{Parameter variation}

\begin{figure*}
  \centering
    \includegraphics[trim = 4mm 23mm 8mm 0, clip, width=0.24 \textwidth]{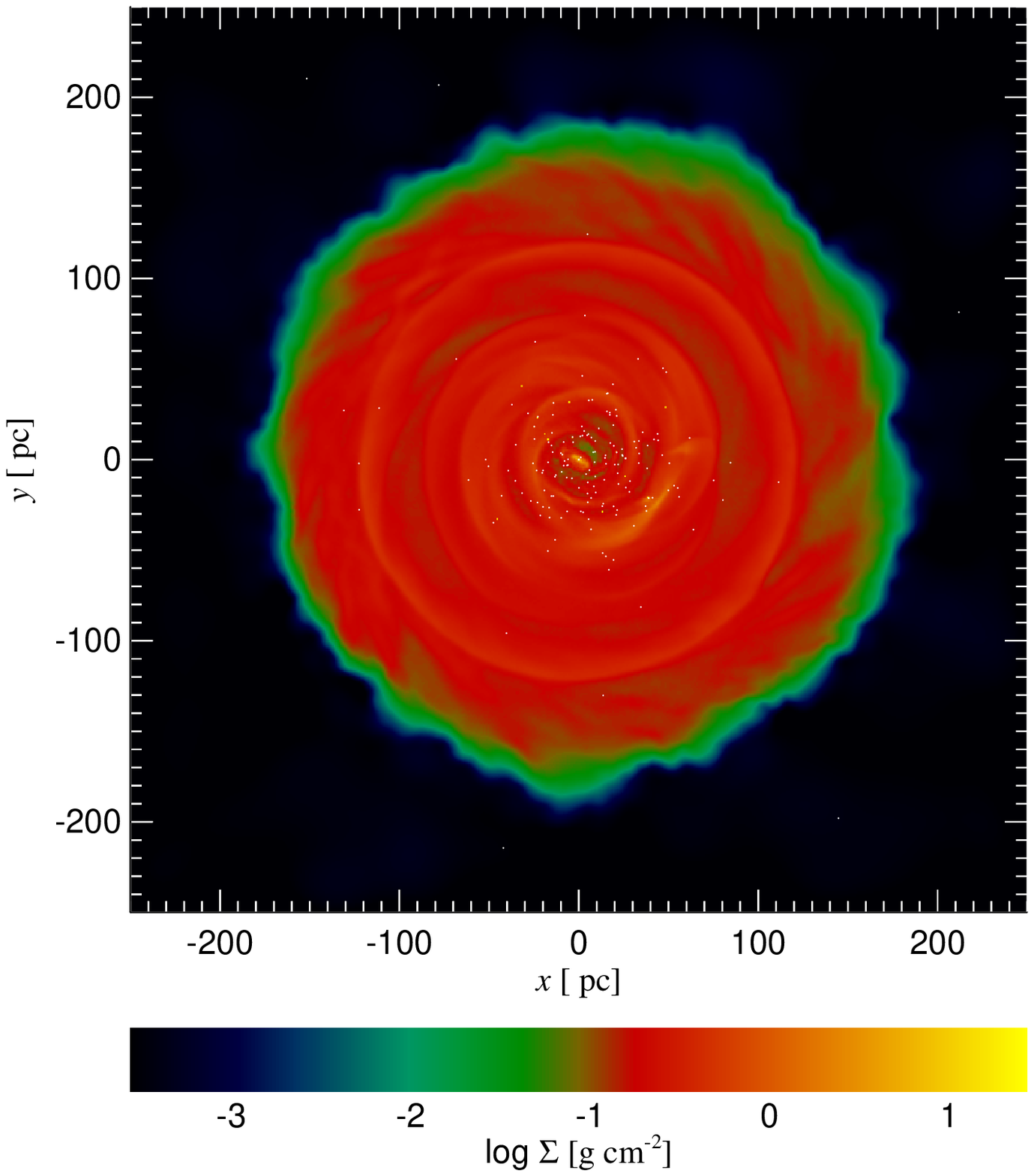}
    \includegraphics[trim = 4mm 23mm 8mm 0, clip, width=0.24 \textwidth]{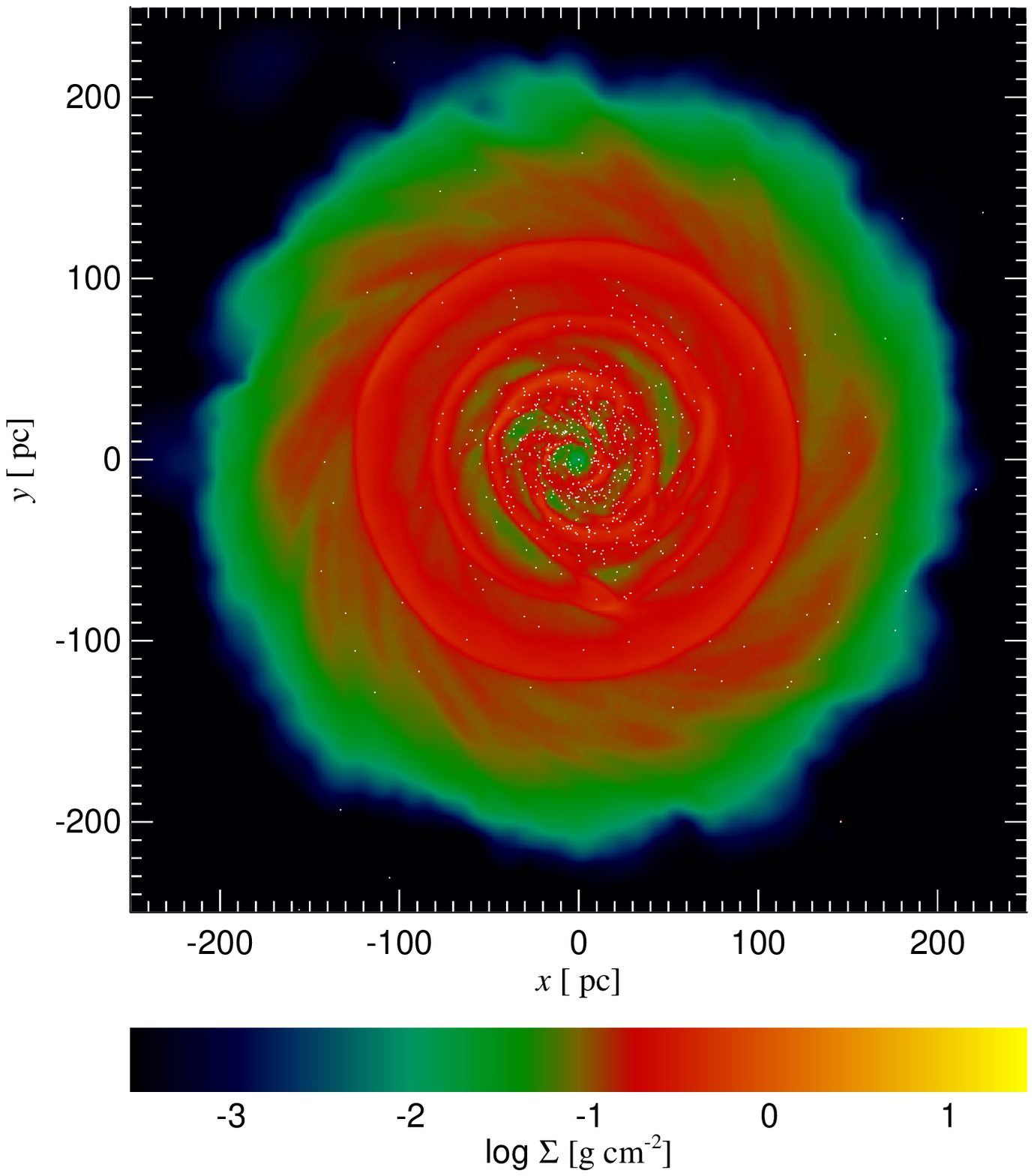}
    \includegraphics[trim = 4mm 23mm 8mm 0, clip, width=0.24 \textwidth]{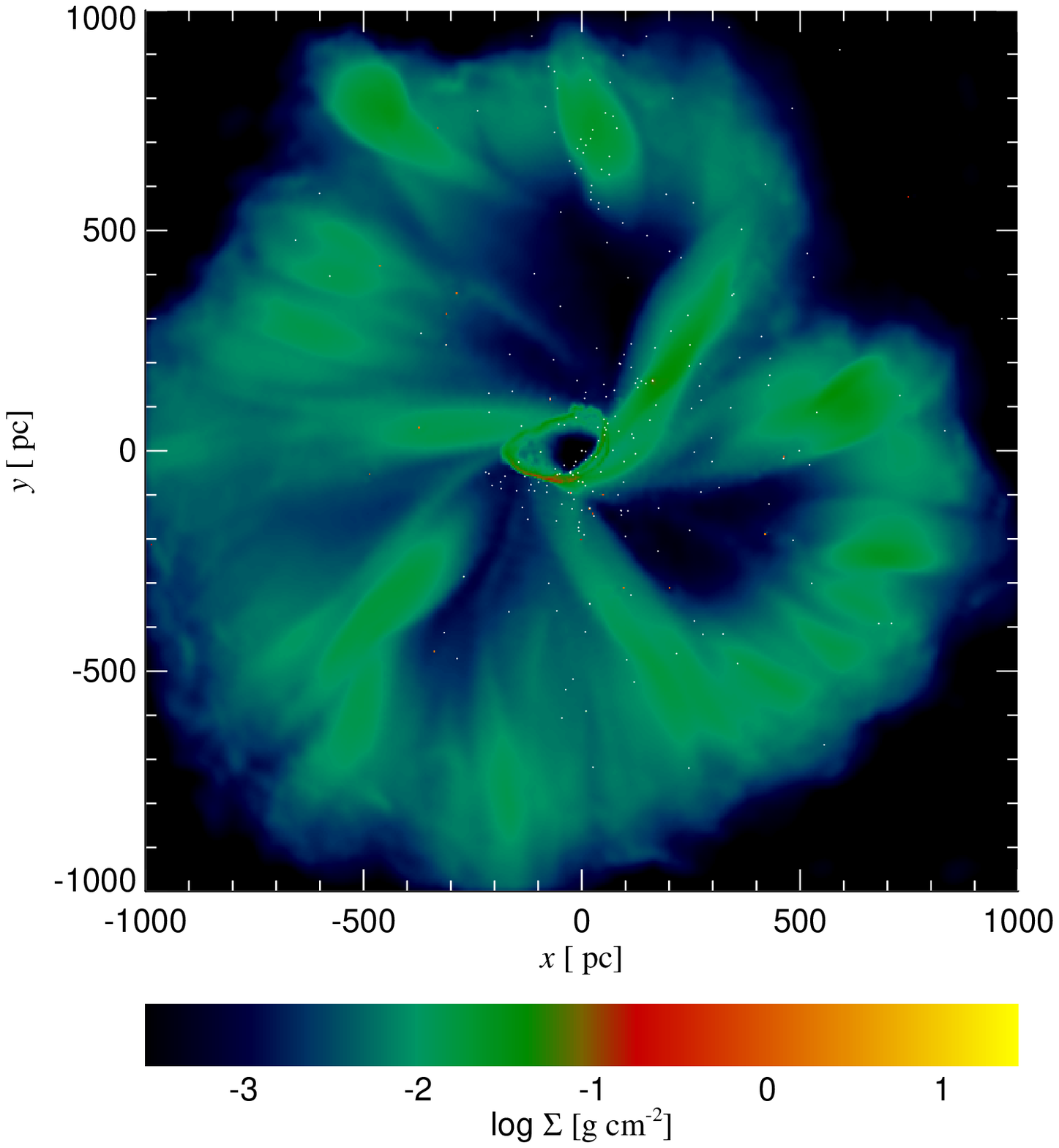}
    \includegraphics[trim = 4mm 23mm 8mm 0, clip, width=0.24 \textwidth]{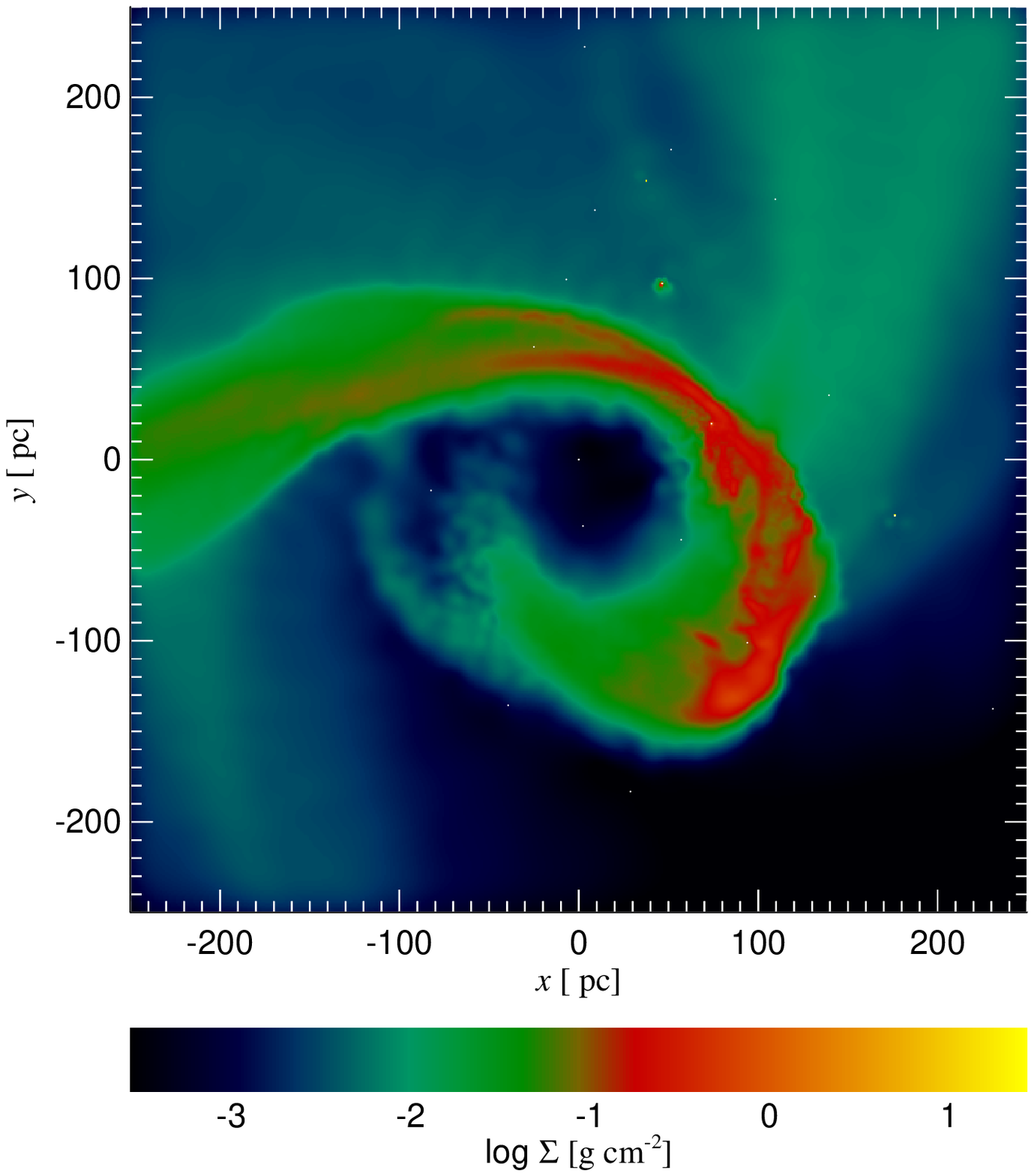}
  \caption{Gas column density of models L0.3 (left panel), T0.3
    (second panel), T1.5 (third panel)  and L0.667T1.5 (last panel) at $t
    = 6$~Myr. The gas morphologies differ significantly, with various
    structures visible. Note the different scale in the third panel.}
  \label{fig:var_morph}
\end{figure*}

\begin{figure}
  \centering
    \includegraphics[width=0.48 \textwidth]{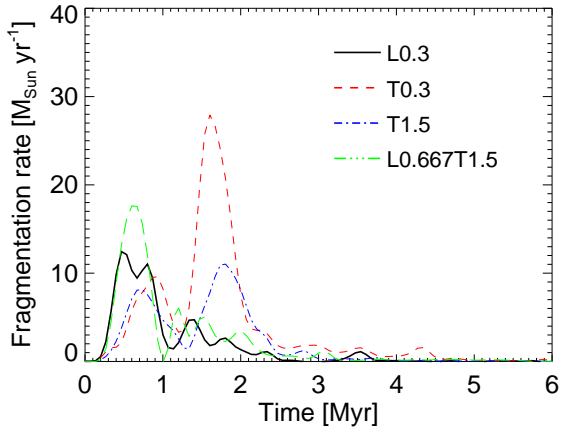}
  \caption{Fragmentation rate in models L0.3 (black solid line), T0.3
    (red dashed), T1.5 (blue dot-dashed) and L0.667T1.5 (green
    long-dashed). In all models, fragmentation is stronger soon after,
    or even during, the AGN phase, gradually decreasing with time as
    the gas distribution relaxes.}
  \label{fig:var_fragrate}
\end{figure}

Figure \ref{fig:var_morph} presents the gas and sink particle
morphology of the final four models with varying parameters. From left
to right, these models are: L0.3, where the AGN luminosity is
decreased to $0.3L_{\rm Edd}$, T0.3, where $t_{\rm q} = 0.3$~Myr,
T1.5, where $t_{\rm q} = 1.5$~Myr, and L0.667T1.5, which has both $L =
0.667L_{\rm Edd}$ and $t_{\rm q} = 1.5$~Myr.

Both L0.3 and T0.3 models show similar morphology. The time evolution
of the models is also similar - a short or weak burst of AGN activity
only affects a small inner part of the CMZ before switching off. The
disturbance propagates outwards, but grows progressively weaker and is
unable to produce gravitational instability in the outer parts of the
CMZ. Some clumps form in the central regions; they produce sink
particles and create spiral waves, which affect the whole disc and are
clearly visible in the morphology plots. The T0.3 model produces more
sink particles (total mass at $t = 6$~Myr is $M_{\rm T0.3} = 2.3
\times 10^7 \; \msun$ and $M_{\rm L0.3} = 1.0 \times 10^7 \; \msun$),
but in both models, essentially all fragmentation occurs at $t <
2$~Myr (see Figure \ref{fig:var_fragrate}, black solid and red dashed
lines). It is interesting that model T0.3 has a single burst of
fragmentation at $t \simeq 1.6$~Myr, which occurs when the perturbed
material falls back to the centre of the disk and reaches highest
density.

In model T1.5, the prolonged action of the AGN unbinds most of the CMZ
gas and allows it to disperse over a very wide region (Figure
\ref{fig:var_morph}, third panel; note that the physical scale is
different than in the other two). This can be expected, because even
though most of the outflow energy escapes in directions perpendicular
to the CMZ, some of that energy is retained and adds to the pure
momentum push in unbinding the gas disc. The gas density in the
central regions is reduced and there is very little gravitational
fragmentation (Figure \ref{fig:var_fragrate}, blue dot-dashed
line). Some spiral perturbations are present, and a ring of relatively
dense gas forms at a distance $d \sim 100$~pc from the centre, but
this ring contains only $\sim 5\times 10^6 \; \msun$ of gas. Sink
particles disperse throughout the volume, with roughly one third of
the total $M_{\rm T1.5} = 1.3 \times 10^7 \; \msun$ escaping from the
central $200$~pc of the Galaxy.

The model L0.667T1.5 provides the same energy input to the CMZ as the
Base models, but spread more gradually. Its evolution is markedly
different from the Base models. A more gradual energy input results in
a much lower density contrast in the CMZ, which then leads to more gas
being expelled from the system and far less fragmentation (Figure
\ref{fig:var_fragrate}, green triple-dot-dashed line). The total sink
particle mass at $t = 6$~Myr is $M_{\rm L0.667T1.5} = 1.3 \times 10^7
\; \msun$. This mass is concentrated in just a few sink particles,
which are distributed with no particular structure throughout the
simulation volume (Figure \ref{fig:var_morph}, right panel).

All four models with parameter variation exhibit very little
fragmentation after $t = 2$~Myr, making the clusters formed in these
simulations older than those observed in the Galactic Centre. In
addition, the morphology of dense gas differs significantly from the
observed ring. These discrepancies suggest that the set of parameters
chosen for the Base and Short models - a $0.7-1$~Myr long
Eddington-limited burst of AGN activity 6 Myr ago - are the best fit
to the \sgra\ activity that could have produced the dense CMZ ring and
massive star clusters.

\subsection{Summary of results}

A brief summary of the main results of the simulations is as follows:

\begin{itemize}

\item A burst of Eddington-limited AGN feedback lasting for
  $0.7-1$~Myr is able to transform an initially smooth gas disc in the
  Galactic centre into a clumpy asymmetric distribution.

\item The morphology of the distribution depends strongly on
  stochastic processes - spiral and gravitational instabilities. It is
  not guaranteed, but seemingly quite likely, that an elliptical ring
  of dense gas forms in the CMZ, and that the major-to-minor axis
  ratio of the ring, as well as the ring size, is very similar to the
  observed one.

\item Fragmentation of the CMZ into sink particles takes off several
  dynamical times after the AGN activity finishes. The dynamical time,
  and hence the age of the sink particle clusters, depends on the
  duration of the AGN episode. For the $t_{rm q} = 1$~Myr episode,
  fragmentation occurs $\sim 3.5-5.5$~Myr after the start of the
  simulation, i.e. $0.5-2.5$~Myr ago. If the AGN episode lasts for
  only $t_{\rm q} = 0.7$~Myr, fragmentation starts earlier, $3.5$~Myr
  ago. The average fragmentation rate during the last 6~Myr is $<10
  \msun$~yr$^{-1}$, with peaks reaching $\ga 30
  \msun$~yr$^{-1}$. Given a typical star formation efficiency on the
  mass scale of the sink particles, $\epsilon_* = 2-10 \%$, the star
  formation rate in the CMZ is $\dot{M}_* \lesssim 0.2-1
  \msun$~yr$^{-1}$, with bursts up to $\sim 3 \msun$~yr$^{-1}$. By $t
  = 6$~Myr, fragmentation in the Short models has dropped to $<2
  \msun$~yr$^{-1}$, corresponding to a star formation rate of $<0.2
  \msun$~yr$^{-1}$; however, fragmentation is still ongoing in some
  Base models.

\item The mass of sink particles formed in the Base simulations is
  $(2.7 - 3.8) \times 10^7 \msun$, predominantly distributed in a few
  clumps, each with mass in the range $(0.5 - 3) \times 10^6
  \msun$. Using the same star formation efficiency as above, the star
  clusters formed should have masses in the range $10^4 - 3\times 10^5
  \msun$. The Short simulations form a slightly smaller total mass of
  sink particles, but the cluster masses are more difficult to
  determine due to numerical effects.

\item An AGN outburst that either has a lower luminosity, $L = 0.3
  L_{\rm Edd}$, or a significantly shorter duration, $t_{\rm q} =
  0.3$~Myr, produces a perturbation to the CMZ that is too weak to
  create a dense ring with embedded young star clusters. Conversely, a
  longer AGN burst duration, $t_{\rm q} = 1.5$~Myr, disrupts the CMZ,
  leaving only a minor fraction of the gas bound in the Galactic
  centre.

\item An AGN outburst with the same total energy output as in the Base
  models, but spread on a longer timescale produces a weaker density
  contrast in the CMZ disc, which leads to an almost non-existent
  dense gas ring and far less sink particle formation. This type of
  outburst might still enhance the existing density contrasts in the
  CMZ, but is unable to shape it to the present state from smooth
  initial conditions.

\end{itemize}

\section{Discussion}\label{sec:discuss}

\subsection{Current properties of the CMZ} \label{sec:current_prop}

The numerical models presented here adequately explain the origin of
some of the peculiar CMZ properties. Notably, an Eddington-limited AGN
episode which started 6 Myr ago and lasted for 0.7-1 Myr could have shaped
an initially axially symmetric CMZ disc into an elliptical, off-centre
dense gas ring with embedded massive molecular clouds and star
clusters. The total mass of dense gas and sink particles (which
represent both stars and gas denser than resolvable in the simulation)
is very similar to the observed value, although this is potentially an
imprint of the initial conditions, and a more (less) massive initial
gas disc would produce a more (less) massive dense gas ring at the
end. The fact that such complex structures can be created from
axisymmetric initial conditions suggests that lopsided nuclear
structures should be a common feature of galaxies (see Section
\ref{sec:rings} below).

The rate of formation of new sink particles at the end of the
simulation, when multiplied by the typical star formation efficiency,
produces a star formation rate estimate similar to the observed value
$\sim 0.06-0.08 \msun$~yr$^{-1}$. This value is at first sight
puzzling on two accounts. It is higher than could be expected from
steady-state arguments \citep{Zubovas2013MNRASb}, i.e. the current SFR
would exhaust the molecular gas reservoir in much less than the age of
the Galaxy. On the other hand, it is several orders of magnitude lower
than expected from the surface density of the gas
\citep{Longmore2013MNRAS}. Explanations for the first mismatch include
bar-induced inflow of material, which replenishes the gas content in
the Galactic centre \citep{Rodriguez2008A&A}, while the second can be
explained by an unusually high level of turbulence in the Galactic
centre molecular clouds, which reduces the star formation rate at a
given surface density \citep{Rathborne2014arXiv}. The AGN-induced
fragmentation model can explain both phenomena simultaneously. The
elevated star formation rate is a transient phenomenon, caused by the
AGN episode compressing the gas, and will likely drop by at least a
factor of a few in the upcoming several Myr. Meanwhile, the molecular
clouds forming from the fragmenting CMZ disc are subject to very high
external pressure from the expanding AGN outflow \citep[the {\it Fermi}
  bubbles; see][]{Zubovas2013MNRASb}, which cause them to have higher
density and turbulence \citep{Elmegreen1997ApJ}, leading to a higher
SFR threshold \citep{Longmore2013MNRAS, Rathborne2014arXiv}.

The AGN-induced fragmentation model naturally explains why the massive
clusters in the CMZ are younger than the nuclear star cluster: the NSC
formed from the gas that fed the AGN outburst, while the CMZ clusters
formed from the perturbed CMZ gas. The ages of the sink particle
clusters forming in the Short models are $\sim 2.5-3.5$~Myr (Figure
\ref{fig:base_fragrate}), consistent with observations of the Arches
and Quintuplet clusters in the Galactic Centre. The precise ages of
star clusters are impossible to determine from the simulations, both
due to insufficient resolution to resolve star formation, and due to
the stochastic evolution of the gas ring, which leads to a variation
in sink particle cluster ages among the simulations with identical
parameters.

\begin{figure}
  \centering
    \includegraphics[width=0.48 \textwidth]{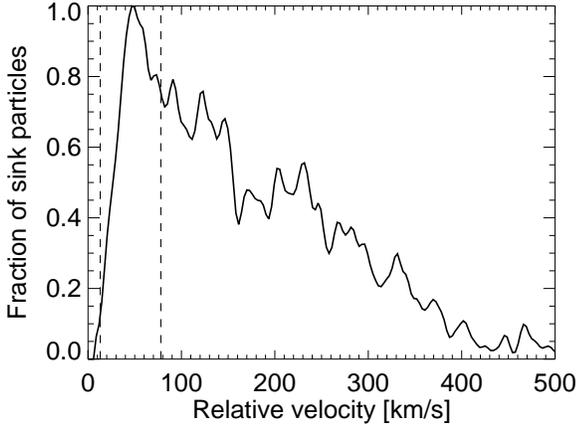}
  \caption{Histogram of relative velocity between sink particles and
    surrounding gas in model Base1. A strong peak around $v_{\rm rel}
    = 45$~km/s is present, with a power-law tail of particles escaping
    from newly-formed clusters. The vertical lines mark the range of
    relative velocities, 13 to 78 km/s, observed by
    \citet{Dong2014arXiv}.}
  \label{fig:base_relvel}
\end{figure}

Recent observations reveal several massive stars in the CMZ which do
not belong to either of the young massive clusters
\citep{Dong2014arXiv}. These stars have velocities relative to their
surrounding gas ranging from $\sim 10$ to $\sim 80$~km/s; some of them
might have formed in-situ, while others might have been ejected from
parent clusters. In my models, both processes occur - some sink
particles form in the dense gas filaments, representing clusters or
associations too small to be resolved in the simulations; other sink
particles form in clusters and then quickly leave them due to
three-body interactions and tidal stripping. In order to better
understand the relative importance of these processes, I plot, in
Figure \ref{fig:base_relvel}, the histogram of relative velocity
between sink particles and surrounding gas (defined as the 40 nearest
SPH particles, with contributions weighted by the SPH kernel) for the
model Base1 at $t = t_{\rm n} = 6$~Myr. A clear peak at $v_{\rm rel} =
50$~km/s is visible, with $\sim 26\%$ of the relative velocities in
the range $\lbrace 13; 78\rbrace$~km/s (vertical dashed lines in the
plot), corresponding to the minimum and maximum relative velocities
observed by \citet{Dong2014arXiv}. There is a large tail of
high-relative-velocity sink particles, indicating that a lot of sink
particles are ejected from clusters, as one would not expect stars
formed in-situ to have relative velocities larger than the background
velocity dispersion $\sigma = 150$~km/s. This may be partly a
numerical effect due to insufficient detail in the star formation
process in the model, as some sink particles might in reality be gas
blobs which escape the clusters, disperse and do not form stars.
Nevertheless, I predict that stars with significantly higher
velocities relative to the surrounding gas should exist in the
Galactic Centre than have been discovered so far.

Certain properties of the CMZ are not reproduced by this model. One
such feature is the twist of the ring \citep{Molinari2011ApJ}, which
is not seen in any of the models presented here. This suggests that
the twist is either an imprint of different initial conditions,
created by a different process than AGN-induced compression and
fragmentation, or a much rarer effect of AGN activity than the effects
that have been seen. A barred potential leads to vertical as well as
radial oscillations of orbits and so could perhaps create the twisted
shape required by the first two explanations \citep{Binney1991MNRAS,
  Hasan1993ApJ}.

Another feature which the models presented here do not capture is the
dynamical properties of the star clusters, i.e. their sizes, density
distributions and mass functions. The lack of these properties is due
to the crude star formation prescription, which only follows the
initial collapse to a self-gravitating high-density core, and due to
insufficient mass resolution in the models. We may expect a high
integrated star formation efficiency in the cores due to the high
external pressure preventing the parent molecular clouds from
dispersing \citep{Elmegreen1997ApJ, Zubovas2014MNRASc} and a slightly
top-heavy IMF due to the aforementioned high pressure
\citep{McKee2002Natur} and the gas dynamics being similar to the
collect-and-collapse model of induced star formation
\citep{Whitworth1994A&A}.

\subsection{Footprint of past \sgra\ activity} \label{sec:footprint}

Comparison of the results of the Base, Short and variation models
allows me to put constraints on the past activity of \sgra. The AGN
episode 6 Myr ago cannot have been much weaker than Eddington-limited
and cannot have lasted much less, or much longer, than 0.7-1
Myr. These results are consistent with the constraints provided by the
existence of the {\it Fermi} bubbles \citep{Zubovas2011MNRAS,
  Zubovas2012MNRASa} and the presence of a 6 Myr old nuclear star
cluster \citep{Paumard2006ApJ}, which is evidence of a star formation
event and a likely AGN feeding event \citep{Hobbs2009MNRAS}. Together,
these pieces of evidence strongly support the argument that an
Eddington-limited AGN episode occured in the Galactic centre $\sim
6$~Myr ago and lasted for $\sim 0.7-1$~Myr. The dynamical footprints
of this evidence can still be seen today, providing a glimpse into the
past state of the Galaxy. Similar evidence in other galaxies may help
investigate their recent evolution (see Section \ref{sec:rings}).

\citet{Sofue2003PASJb} suggested that star formation in the Galactic
centre might proceed in recurrent episodes, with footprints of many
such episodes visible throughout the Galaxy. In his model, the North
Polar Spur was formed from a starburst that occurred 15 Myr ago, and
the Galactic Centre Lobe is the remnant of a several-Myr old
starburst. My results are consistent with this picture. A burst of AGN
activity shuts off the feeding reservoir for a period of several Myr,
but allows for more feeding later (see Section \ref{sec:fuelling}
below); this process may cause AGN luminosity to fluctuate in short
bright bursts separated by periods of insignificant activity. Since
each AGN episode may lead to compression of surrounding gas and a
period of enhanced star formation, the star formation history in the
Galactic centre should be composed of short bursts separated by longer
periods of inactivity, exactly as suggested by \citet{Sofue2003PASJb}.

\subsection{General properties nuclear rings affected by AGN feedback} \label{sec:rings}

Many galactic centres contain large amounts of dense gas arranged in
rings or discs \citep{Kormendy2004ARA&A, Knapen2005A&A, Krips2007A&A,
  Boeker2008AJ}. This feature is often explained as a consequence of
bar-driven inflow of material from kpc scales \citep{Combes1985A&A,
  Yuan1997ApJ, Sakamoto1999ApJ, Combes2003ASPC, Kormendy2004ARA&A,
  Sheth2005ApJ, Rodriguez2008A&A}, although there are unbarred
galaxies with nuclear stellar rings \citep{Knapen2004A&A,
  Silchenko2006AJ}. A bar-driven gas inflow process predicts several
morphological features of the rings: ellipticity \citep[since the
  rings follow triaxial $x_2$ orbits;][]{Morris1996ARA&A,
  Knapen2002MNRAS, Kim2011ApJb}, more massive gas clumps close to the
orbit crossing \citep[since material moves slowest
  there;][]{Benedict1996AJ} and an azimuthal gradient of star cluster
ages \citep[since star formation is triggered at the gas injection
  points;][]{Falcon-Barroso2007Msngr, Boeker2008AJ}. All three
properties have been observed in some galaxies, but are not ubiquitous
\citep{Knapen2005A&A, Falcon-Barroso2007Msngr, Boeker2008AJ}.

The model of AGN-induced compression of the CMZ gas presented in this
paper might appear as an alternative to the bar-driven inflow
model. However, it is more appropriate to view the two models as
complementary. The transport of gas into the central parts of the
galaxy is mediated by bar potential or caused by minor
mergers. Afterward, the nuclear gas ring is affected by both the
aspherical gravitational potential and episodic AGN feedback. The
latter has several important effects on such nuclear gas rings:
facilitating the formation of more chaotic ring morphology, causing
bursty and elevated star formation, and producing a correlation
between star cluster ages and ring sizes.

\subsubsection{Ring morphology}

An AGN outflow disturbs the gas ring radially and induces
instabilities. These alter the shape of the ring, making it irregular,
rather than axially symmetric. This effect has been observed both
directly, in NGC 613, where AGN activity is associated with
disturbances in gas morphology \citep{Falcon-Barroso2014MNRAS}, and
can be inferred for many other galaxies, where irregular morphologies
of gas rings are common \citep{Sarzi2007MNRAS, Boeker2008AJ}.

The location of most massive molecular clouds is also random in the
AGN-induced compression model. In our Galactic Centre, the star
forming regions Sgr B and Sgr C are located close to the ends of the
CMZ ellipse \citep{Molinari2011ApJ}, as predicted by a bar-driven
inflow model. However, other galaxies do not show such regular
structures \citep{Knapen2005A&A, Sarzi2007MNRAS,
  Falcon-Barroso2014MNRAS}, so a more stochastic process of generating
molecular clouds is preferred there.

Finally, AGN-induced fragmentation also leads to random locations of
star formation and hence to random azimuthal distribution of star
cluster ages. This is called a ``popcorn model'', as opposed to the
``pearls on a string'' model predicted by bar-driven inflows
\citep{Boeker2008AJ}. In this latter model, star and cluster formation
occurs predominantly at the ends of the ellipsoidal ring, where the
gas from the galactic bar enters the ring; the young clusters then
travel along the ring, producing an azimuthal gradient of stellar
ages, with the youngest stars at the ends of the ellipse and oldest
stars just upstream of those points. Many galaxies show azimuthal age
gradients of star clusters in nuclear rings, but this isn't a
ubiquitous feature \citep{Sarzi2007MNRAS, Boeker2008AJ}.

\subsubsection{Star formation rates}

The mass flow rate of a bar-driven gas inflow varies little over time,
since the stellar bar is typically long-lived \citep{Kim2011ApJb,
  Seo2013ApJ}.  This results in an approximately constant star
formation rate and predicts a rather uniform spread of star cluster
ages. Observations, however, reveal that star formation in rings is
more likely to be bursty \citep{Allard2006MNRAS, Sarzi2007MNRAS},
suggesting a need for episodic perturbations. AGN feedback can provide
just such perturbations, increasing the SFR over the long-term
average. AGN feedback also explains why the observed star formation
efficiency in nuclear starbursts is much higher than in typical
star-forming regions \citep{Kennicutt1998ApJ}: AGN outflows create
conditions of very high pressure around the gas ring, which prevent
gas from escaping and enhance star formation \citep{Elmegreen1997ApJ,
  Zubovas2014MNRASc}.

\subsubsection{Ages of star clusters}

The sequence of events at the basis of my model - AGN fuelling and
outburst, mixing of gas in the nuclear ring and formation of molecular
clouds and star clusters - predicts a clear difference between the age
of star clusters in the nuclear ring and the time since the last
significant SMBH accretion episode, which should coincide with the
last star formation episode in the central few parsecs of the
galaxy. Therefore, the age difference between the youngest stellar
population in the nuclear star cluster (NSC) and the youngest star
clusters in the nuclear gas ring can be expressed as
\begin{equation} \label{eq:tage}
\Delta t_{\rm age} = K t_{\rm dyn} \left(R_{\rm ring}\right) \propto
R_{\rm ring} / \sigma_{\rm gal},
\end{equation}
where $R_{\rm ring}$ is the radius of the nuclear gas ring and
$\sigma_{\rm gal}$ is the velocity dispersion in the galaxy spheroid.
$K$ is a constant, here assumed to be independent of ring properties
or the larger galactic environment; this assumption is supported by
the results of the Base and Short models (see Section
\ref{sec:short}). Using the parameters of the Milky Way, I find $K
\sim 3$. Current observations cannot easily distinguish the youngest
populations in NSCs, since those populations probably comprise only a
small fraction of the total NSC mass (in our Galactic centre, the mass
of stars formed in the burst $6$~Myr ago is a few per cent of the
total NSC mass), but in the future, it should be possible to test this
correlation.

One example where this correlation can be seen is the galaxy NGC 1614,
which hosts a nuclear starburst \citep{Alonso-Herrero2001ApJ,
  Koenig2013A&A}. The galaxy contains a very young NSC, $t_{\rm age,
  NSC} = \left(7 \pm 1\right)\times10^6$~yr \citep{DeGrijs2005MNRAS};
its central velocity dispersion is $\sigma_{\rm gal} = 164 \pm 8$~km/s
\citep{Hinz2006ApJ}. The nuclear gas ring extends radially from $100$
to $350$~pc \citep{Xu2014arXiv}, with most star-forming clumps found
at a radius $R_{\rm ring} = 230$~pc. Using equation (\ref{eq:tage})
above and $K = 3$ gives $\Delta t_{\rm age, NGC 1614} = 4.2$~Myr,
i.e. the star clusters in the nuclear gas ring should be $\sim
2.8$~Myr old. The precise ages of the star clusters in the nuclear
ring are not known, but the presence of large HII regions and evidence
of ongoing star formation suggests that they are younger than $5$~Myr
\citep{Alonso-Herrero2002AJ}, consistent with the model prediction. In
fact, there is evidence that the nuclear starburst in NGC 1614 is
propagating outward \citep{Alonso-Herrero2001ApJ}, which may simply be
a result of longer dynamical time, and hence slower evolution of a
perturbed gas ring, at larger radii. This explanation predicts an
apparent outward propagation of the starburst with velocity $v_{\rm
  out} = \sigma_{\rm gal} / K \sim 55$~km/s, in excellent agreement
with observations \citep[$v_{\rm out, obs} =
  60$~km/s][]{Alonso-Herrero2001ApJ}.

The fact that a burst of nuclear star formation follows the AGN
activity episode with a delay gives an opportunity to put constraints
on possible past AGN activity in a galaxy. This was discussed for our
Galaxy in Section \ref{sec:footprint} above. An interesting example is
M82, the prototypical starburst galaxy. It contains a large number of
young star clusters in its $\sim 250$~pc core, with ages $\sim 10$~Myr
\citep{Satyapal1997ApJ, Melo2005ApJ}. The velocity dispersion in the
bulge of M82 is $\sigma \simeq 100 \pm 20$~km/s
\citep{Gaffney1993ApJ}. Assuming that the core is a ring-like
structure, the time since the most recent AGN burst should be
$17.5$~Myr or so - long enough for the AGN to have switched off and
its most direct effects to have faded, but short enough for an outflow
to remain visible \citep{King2011MNRAS}. A part of the outflow
observed in M82 may be a fossil AGN outflow, rather than purely due to
starburst activity. On the other hand, if one assumes that the star
formation propagates inwards from the larger-scale older starburst
\citep{DeGrijs2001A&G}, the time delay between the young star cluster
ages and AGN outburst implies a powerful AGN activity episode starting
$5\pm2.5$~Myr ago; one might expect obvious evidence of recent
powerful AGN activity in this case, which is not detected \citep[only
  a low-luminosity AGN is detected in M82;][]{Matsumoto1999PASJ}. The
outward propagation of the starburst is also supported by observations
of young star cluster ages, giving a propagation velocity $\sim
50$~km/s \citep{Satyapal1997ApJ}, qualitatively consistent with the
model of an AGN-induced starburst.

\subsection{AGN fuelling} \label{sec:fuelling}

Although this is not the primary goal of the paper, the results
presented herein offer several suggestions regarding fuelling of AGN
in the presence of feedback. 

The usual mechanism invoked for transporting gas from outer regions of
galaxies toward the centre is gravitational torques from stellar bars
\citep[also see Section \ref{sec:rings} above]{Combes2003ASPC}.
However, they tend to cause gas to accumulate in nuclear rings several
hundred pc in size. Further gas infall is thwarted by the bar
resonances and change of orbit types \citep{Regan2003ApJ,
  Garcia-Burillo2005A&A}. One possible mechanism for bringing gas
closer to the AGN is by action of nested bars
\citep{Shlosman1989Natur, Maciejewski2002MNRAS}, but only a few
galaxies show evidence for such structures. Other explanations invoke
transient effects, such as increased gas turbulence due to starburst
activity in the central kiloparsec, which enhances mixing of gas of
different angular momentum and allows gas to penetrate further into
the galactic centre \citep{Hobbs2011MNRAS}.

My results show that mixing of gas angular momentum can be achieved by
the AGN itself, by means of feedback acting upon gas of spatially
varying angular momentum (see Figure \ref{fig:angmom} and Section
\ref{sec:mixing}). In the case of the CMZ, this mixing is not very
efficient, because gas is already in ordered motion around the centre,
i.e. the mean angular momentum of the gas is large and the formal
circularization radius of the whole CMZ is also large. This allows
only a small fraction of all the gas to lose most of its angular
momentum via the action of local gravitational torques produced by
gravitational instabilities in the CMZ. In a more general case, where
gas is falling in through the bulge, the motion is generally less
ordered and the total angular momentum of the gas is small. In that
case, mixing of gas, enhanced by AGN feedback, can cause gas clouds
with very different angular momenta to collide and fall toward the
centre. This scenario was investigated analytically by
\citet{Dehnen2013ApJ}, who found that SMBHs can easily be fed at their
Eddington limit by repeated short bursts of feedback. The results of
my models, showing efficient angular momentum mixing, are consistent
with the basic picture of this work. In particular, this irregular
mixing might lead to massive reservoirs of gas accumulating very close
to the SMBH, within the sphere of influence. This might be the case in
Andromeda, where the double nucleus shows a massive reservoir of gas
within a few pc of the SMBH, but with little star formation
\citep{Tremaine1995AJ, Bender2005ApJ}.

\subsection{Outstanding issues}

The study presented here is a first step in understanding the effects
of AGN activity upon nuclear gas rings. As a result, a number of
simplifying assumptions had to be made in order to isolate the salient
effects. Below, I discuss the importance of the effects neglected by
making these assumptions.

\subsubsection{Gas equation of state} \label{sec:eos}

The models presented in this paper adopt a realistic gas
heating-cooling prescription above $T = 10^4$~K, but prevent gas from
cooling below this value. This is done because cooling above $10^4$~K
is less dependent upon detailed effects of gas self-shielding, which
my models cannot resolve. In principle, the temperature floor
stabilizes the disc against gravitational instabilities and washes out
possible small-scale structures in the perturbed gas rings, leading to
a lower fragmentation rate than might be expected if the gas was
allowed to cool to lower temperatures. Since the density threshold for
fragmentation varies as $n_{\rm J} \propto T^3$
(eq. \ref{eq:rhojeans}), a temperature floor at $T = 100$~K would
reduce this threshold by a factor $10^6$. However, the peaks in the
gas density distribution are extremely sharp and tall, rising more
than 6 orders of magnitude above the typical densities in the gas
disc/ring. Therefore, even such a significant change in the
temperature floor should not affect the fragmentation rate
significantly.

Feedback from star formation might also affect the fragmentation rate,
since stars can heat their surroundings and hamper or shut off further
fragmentation in their vicinity \citep{Bate2009MNRAS}). Proper
implementation of feedback, however, requires much higher mass
resolution than in the simulations presented here, sufficient to
resolve massive stars and small stellar associations.

\subsubsection{Gravitational potential}

The set up of the models presented in this paper makes an assumption
of a spherically symmetric background potential. The presence of a
Galactic bar \citep{Bissantz2002MNRAS, Babusiaux2005MNRAS} makes the
actual potential triaxial, which may significantly affect the results
of these simulations.  Qualitatively, an asymmetric potential should
enhance the asymmetry in the final gas distribution, so the main
findings of this paper should still be valid.

\subsubsection{Initial conditions}

The initial gas distribution used in the simulations presented here is
necessarily strongly idealised. One reason for choosing a smooth disc,
rather than a realistic lumpy distribution, is that any enhancement of
gas fragmentation is likely to be even stronger when more realistic
gas distributions are considered. This would occur because a realistic
gas distribution has pre-existing clumps and density inhomogeneities,
which would be enhanced by a passing AGN outflow. On the other hand,
some parts of the disc may be cleared, rather than compressed, so the
induced star formation rate would be smaller in a more realistic
situation.

Given that most of the outflow impacting the CMZ is deflected
perpendicularly, it might seem that the outflow has only a weak effect
on the Galaxy on kpc scales. This was found in some previous
simulations of high-redshift, therefore gas-rich galaxies
\citep{Gabor2014MNRAS}. However, one has to keep in mind that even
though the outflow is collimated, its opening angle is still large, of
order $\pi/2$ for each cone \citep{Zubovas2012MNRASa}. This value is
consistent with recent observational results
\citep{Nardini2015Sci}. Therefore, a large part of the galaxy is
affected by the outflow directly (radially), and the rest of the gas
may be compressed perpendicularly \citep{Silk2005MNRAS,
  Gaibler2012MNRAS, Zubovas2013MNRASb}. If the initial conditions are
made more realistic, the outflow might escape in even more directions,
leading to an even larger effect on the host galaxy.

\subsubsection{Numerical resolution}

Another important aspect of the simulation is the numerical
resolution. With the resolution chosen for the simulations, the
smoothing length of CMZ gas particles, $h_{\rm SPH}$, is between $0.1
- 2$ times the scale height of the disc at the beggining of the
simulation, with values $h_{\rm SPH} > H_{\rm CMZ}$ reached only on
the surface of the disc. This means that the vertical thickness of the
CMZ disc is typically resolved with a few tens of particles, making
the vertical structure marginally resolved, and hence fragmentation in
the disc should not be artificially suppressed
\citep{Nelson2006MNRAS}.

The masses of fragments forming in the CMZ depend only weakly on
numerical resolution. While initially they form from single SPH
particles, they quickly accrete and/or merge with other particles,
growing in mass to several hundred time the SPH particle
mass. Therefore, the masses of those fragments are generally resolved,
from a purely numerical perspective. Their masses would undoubtedly be
affected if additional physics (such as stellar feedback) is added to
the simulation.

Numerical resolution also affects the interaction between the AGN wind
and the CMZ gas. The momentum and energy carried by each virtual
particle is scaled to the total SPH particle number, so this direct
interaction should not be strongly affected. However, the effective
boundary of the CMZ disc, where gas can absorb the wind energy and
evaporate vertically, is always $\sim h_{\rm CMZ}$ thick, since
virtual particles are typically absorbed in a region of this length
scale. Therefore, with higher resolution, this boundary becomes
physically thinner and less dense. This leads to less evaporation from
the disc, but more importantly, the gas is heated to higher
temperatures by the interaction with the AGN wind. Therefore, it cools
more slowly (since $t_{\rm cool} \propto T^{1/2}$ for Bremsstrahlung
cooling, dominant at the large temperatures relevant in this case) and
both evaporates and passes energy to the surrounding matter more
rapidly. As a result, the disc might expand more both in the radial
and vertical directions. Given that this effect depends not just on
resolution, but also on the gas heating-cooling prescription, a
thorough analysis is beyond the scope of this paper.

\section{Conclusion}\label{sec:concl}

I presented the results of hydrodynamical simulations of a brief AGN
activity episode and the associated outflow affecting the gas disc
around the Galactic centre. The simulation set up is designed to mimic
the properties of our Galaxy, and the time since the AGN outburst,
$6$~Myr, is fixed by the age of the young stars in the central
parsec. The initial distribution of the gas has the size and mass of
the Central Molecular Zone, but is distributed in a smooth disc. 

With this setup, I show that an Eddington-limited AGN outburst lasting
for $0.7-1$~Myr can by itself transform the smooth gas disc into a
narrow, elliptical and off-centre ring of dense gas clumps and young
star clusters. The masses of star-forming regions and star clusters
agree rather well with the observed features in the CMZ - the Sgr B2
molecular cloud and the Arches and Quintuplet star clusters - and the
shape of the ring is also similar to observations. However, the
stochastic nature of gravitational instabilities, which cause the
formation of structures, means that there is a lot of variation among
the models and more concrete predictions are difficult to make.

Several properties of the CMZ cannot be explained with this
model. Most importantly, the CMZ is known to be twisted
perpendicularly to its plane, a feature not reproduced in any of my
simulations. This twist may potentially be explained by the action of
an asymmetric background potential, i.e. the Milky Way bar, but
further work is required to test this hypothesis. Certain other
observed features of the CMZ are consistent with a purely bar-driven
gas inflow model. A recent AGN outburst and associated outflow should
be considered as an addition to this model, not a replacement. In the
future, I plan to model the response of a more realistic barred
galactic environment to an AGN outflow.

Given that past AGN activity can explain many features of the CMZ,
these features can be used to constrain the duration and luminosity of
the AGN episode. These constraints agree with previous constraints
based on the {\it Fermi} bubbles \citep{Zubovas2011MNRAS,
  Zubovas2012MNRASa}. Together, these aspects are dynamical footprints
of past AGN activity, and point to the possibility of investigating
activity history in other galaxies as well, by finding appropriate
dynamical structure that might have been affected by past AGN
outflows.

Nuclear gas rings and star formation within them is a common feature
in many galaxies. While bar-driven gas inflows can explain many
features of these rings, some aspects, such as high star formation
efficiencies during repeated starbursts, disturbed morphologies and
radial gradients of star cluster ages, can be more easily explained by
the action of AGN outbursts. Overall, AGN outbursts have a
significant influence upon the properties of host galaxy centres, and
future observations will help quantify this influence.

\section*{Acknowledgments}

I thank Vladas Vansevi\v{c}ius, Donatas Narbutis and Sergei Nayakshin
for useful discussions during the preparation of this manuscript, and
the anonymous referee for a thorough and helpful report. This research
is funded by the Research Council of Lithuania grant
no. MIP-062/2013. Some numerical simulations presented in this work
were performed on resources at the High Performance Computing Center
¥HPC Sauletekis´ in Vilnius University Faculty of Physics.

\end{document}